\documentclass[twocolumn,journal, bookmarks=false]{IEEEtran}
\IEEEoverridecommandlockouts
\usepackage{verbatim}
\usepackage{siunitx}
\usepackage{color}
\usepackage{graphicx}
\usepackage{amsmath}
\usepackage{amssymb}
\usepackage{algorithm}
\usepackage{algorithmic}
\usepackage{amsmath}
\usepackage{multirow}
\usepackage{booktabs}
\usepackage{array}
\usepackage{amsthm}
\usepackage{stfloats}
\usepackage{caption}
\usepackage{subfigure}
\usepackage{bm}
\usepackage{epstopdf}
\usepackage{setspace}
\usepackage{gensymb}
\usepackage{url}
\usepackage{verbatim}   
\usepackage{balance}
\usepackage{cite}

\usepackage{hyperref}
\usepackage{cleveref} 
\hypersetup{colorlinks = true,
    linkcolor = black,
    urlcolor = black,
    citecolor = black}
\newtheorem{prop}{Proposition}

\usepackage{hhline}

\allowdisplaybreaks[4]

\title{OFDM-ISAC Beyond CP Limit: Performance Analysis and Mitigation Algorithms
\thanks{This work was supported in part by the National Natural Science Foundation of China (Grant No. 62371090, 62471086, and U25A20393) and in part by the U.S. National Science Foundation under Grant CCF-2322191. The work of Rang Liu and A. Lee Swindlehurst was supported by the U.S. National Science Foundation under grant CCF-2322191. The associate editor coordinating the review of this article and approving it for publication was Prof. Rodrigo de Lamare. \textit{(Corresponding author: Ming Li.)}}
\thanks{P. Li and M. Li are with the School of Information and Communication Engineering, Dalian University of Technology, Dalian 116024, China (e-mail: lipeishi@mail.dlut.edu.cn; mli@dlut.edu.cn).}
    \thanks{R. Liu and A. Lee Swindlehurst are with the Nhu Department of Electrical Engineering and Computer Science, University of California, Irvine, CA 92697, USA (e-mail: rangl2@uci.edu; swindle@uci.edu).}
    \thanks{Q. Liu is with the School of Computer Science and Technology, Dalian University of Technology, Dalian 116024, China (e-mail: qianliu@dlut.edu.cn).}
}

\author{Peishi Li,
    Ming Li,~\IEEEmembership{Senior Member,~IEEE,}
    Rang Liu,~\IEEEmembership{Member,~IEEE,}  
    Qian Liu,~\IEEEmembership{Member,~IEEE,} \\
    and A. Lee Swindlehurst, ~\IEEEmembership{Life Fellow,~IEEE}}

\begin{document}

\maketitle
\thispagestyle{empty}

\begin{abstract}
    Orthogonal frequency division multiplexing (OFDM) is well-suited for integrated sensing and communications (ISAC), yet its cyclic prefix (CP) is dimensioned for communications-grade multipath and is generally insufficient for sensing. When echoes exceed the CP duration, inter-symbol and inter-carrier interference (ISI/ICI) break subcarrier orthogonality and degrade sensing. This paper presents a unified analytical and algorithmic framework for OFDM-ISAC beyond the CP limit. We first develop a general echo model that explicitly captures the structured coupling of ISI and ICI caused by CP insufficiency. Building on this model, we derive closed-form signal-to-interference-plus-noise ratio (SINR) and range-Doppler Map (RDM) second-order moment, together with an approximate peak sidelobe level ratio (PSLR), both of which are shown to deteriorate approximately linearly with the normalized excess delay beyond the CP. To mitigate these effects, we propose two standard-compatible successive interference cancellation (SIC) methods: SIC-DFT, a low-complexity DFT-based scheme, and SIC-ESPRIT, a super-resolution subspace approach. Simulations corroborate the analysis and demonstrate consistent gains over representative benchmarks. Both algorithms provide more than $4$~dB SINR improvement under CP-insufficient conditions, while SIC-ESPRIT reduces range/velocity root-mean-square-errors (RMSE) by about one order of magnitude, approaching the performance achievable with a sufficiently long CP. These results offer both theoretical insight and practical solutions for reliable long-range OFDM-ISAC sensing beyond the CP limit.
\end{abstract}


\section{Introduction}
Integrated sensing and communication (ISAC) has emerged as a pivotal enabling technology for next-generation wireless systems, transforming traditionally isolated radar and communication networks into unified, multifunctional platforms \cite{Zhang_VTM_2021}-\cite{LiuR_PROCIEEE_2026}. By jointly leveraging spectral resources, hardware architectures, and advanced signal processing techniques, ISAC achieves substantial improvements in spectrum efficiency and system integration, enabling emerging applications such as autonomous driving, smart manufacturing, and environment-aware wireless connectivity \cite{Zeng_WCM_2021, LiuR_WC_2025}. Among various waveform candidates, orthogonal frequency division multiplexing (OFDM) has attracted particular attention owing to its widespread adoption in contemporary wireless standards (e.g., 5G NR, Wi-Fi 6/7) and its inherent suitability for ISAC applications, such as efficient decoupled estimation of delay and Doppler \cite{Braun_RadarConf_2010, Sturm_Proc_2011}, flexible time-frequency resource allocation \cite{Keskin_TSP_2021, Li_TSP_2025_resource}, and low range sidelobes \cite{Liu_TSP_2025, Liu_TIT_2025}.

A fundamental constraint in OFDM-based ISAC systems stems from the cyclic prefix (CP), originally introduced in communication systems to combat inter-symbol interference (ISI) caused by multipath propagation. For communication systems, CP durations are designed based on microsecond-level delay spreads, typically adequate for wireless channels. \textcolor{black}{For OFDM radar sensing with standard fixed-window demodulation, CP-induced ISI/inter-carrier interference (ICI)-free processing requires the two-way echo delay to fall within the CP, so that the circular convolution structure is preserved over the FFT window. For instance, considering the 5G NR normal CP configuration with a subcarrier spacing of $120$~kHz, the standard CP duration is only $0.59~\mu$s, theoretically restricting the interference-free sensing range to roughly $90$~m \cite{3gpp_standards}. The resulting CP-protected sensing range is far less than the unambiguous range and is usually insufficient for outdoor sensing applications.} Thus, practical ISAC systems must reliably operate \emph{beyond the CP limit}, where sensing requirements fundamentally diverge from communication specifications.

Several studies have recently investigated the impact of insufficient CP on OFDM sensing performance \cite{Wang_WiOpt_2023,Wang_TVT_2025,Zhou_PIMRC_2024,Li_Globecom_2025}. These analyses reveal that when target echoes extend beyond the CP duration, significant ISI and ICI arise, causing elevated sidelobe levels in the range-Doppler map (RDM) and substantially degraded parameter estimation accuracy \cite{Wang_WiOpt_2023, Wang_TVT_2025, Zhou_PIMRC_2024}. Unlike communication systems, where moderate ISI can typically be mitigated through equalization, radar sensing relies heavily on coherent signal accumulation across subcarriers and symbols, making it inherently more sensitive to structured interference. Existing theoretical analyses, however, often simplify ISI/ICI modeling by assuming statistical independence from interference-free components or by adopting Gaussian approximations. These simplifications limit analytical precision and fail to capture the intrinsic structured coupling of ISI/ICI components in OFDM sensing scenarios. Consequently, a rigorous and accurate analytical performance characterization for OFDM-ISAC systems beyond the CP limit remains largely unexplored.

Aside from performance characterization, recent works have also examined methods for mitigating the detrimental effects of insufficient CP. \textcolor{black}{Existing beyond-CP OFDM sensing methods can be broadly grouped into waveform design, sliding-window receiver processing, and coherent compensation, as summarized in Table~\ref{Tab:literature_comparison}.
    On the transmitter side, CP structure modification \cite{Zhou_PIMRC_2024, Yuan_CL_2023, Li_ICCC_2024}, pilot pattern redesign \cite{Tang_CL_2025}, and symbol-level waveform optimization \cite{Jiang_WCL_2025} have been proposed to suppress beyond-CP interference. Although effective, these methods often compromise spectral efficiency and deviate from standardized OFDM signaling. On the receiver side, sliding-window or delay-dependent processing preserves the OFDM transmit waveform but no longer relies on a single fixed standard FFT demodulation window at the sensing receiver. By shifting the receive window or testing multiple FFT-demodulation hypotheses, these methods can obtain favorable observations over different delay regions \cite{Ozturk_MILCOM_2025,Xu_TSP_2025}. However, they require additional delay-window searches or selections and do not explicitly characterize or cancel the structured ISI/ICI observed under standard fixed-window demodulation. Other receiver-side methods retain the standard demodulation window and perform coherent compensation, including time-domain coherent compensation (TDCC) \cite{Wang_WiOpt_2023, Wang_TVT_2025}, virtual-CP reconstruction \cite{Wu_JSAC_2022}, frequency-domain coherent compensation (FDCC) \cite{Geiger_SCC_2025}, and multi-target coherent compensation (MTCC) \cite{Geiger_SCC_2025}. These methods enhance the useful echo component under fixed-window demodulation, but residual structured ISI/ICI remains and may be amplified together with noise.} Therefore, explicit modeling and mitigation of the structured ISI/ICI under a standard OFDM signaling and receiver processing remain insufficiently explored in the existing literature.

\begin{table*}[!t]
    \begingroup
    \footnotesize
    \centering
    \caption{\textcolor{black}{Representative Beyond-CP OFDM Sensing Approaches}} \label{Tab:literature_comparison}
    \textcolor{black}{
        \begin{tabular}{>{\raggedright\arraybackslash}p{0.25\textwidth}
            >{\raggedright\arraybackslash}p{0.34\textwidth}
            >{\raggedright\arraybackslash}p{0.35\textwidth}}
            \toprule
            \textbf{Representative works} & \textbf{Key idea}                                     & \textbf{Main limitation / requirement} \\
            \midrule
            \textbf{Waveform design} \cite{Zhou_PIMRC_2024,Yuan_CL_2023,Li_ICCC_2024,Tang_CL_2025,Jiang_WCL_2025}
                                          & Modify CP/pilots/symbols.
                                          & Spectral-efficiency loss or nonstandard signaling.                                             \\
            \addlinespace[4pt]
            \textbf{Sliding-window processing} \cite{Ozturk_MILCOM_2025,Xu_TSP_2025}
                                          & Shift/test sensing FFT windows.
                                          & Delay-window search; nonstandard receiver processing.                                          \\
            \addlinespace[4pt]
            \textbf{Coherent compensation} \cite{Wang_WiOpt_2023,Wang_TVT_2025,Wu_JSAC_2022,Geiger_SCC_2025}
                                          & Compensate useful echoes in the fixed window.
                                          & Residual structured ISI/ICI and noise amplification.                                           \\
            \addlinespace[4pt]
            \textbf{Proposed SIC framework}
                                          & Analyze and cancel fixed-window structured ISI/ICI.
                                          & Possible SIC error propagation.                                                                \\
            \bottomrule
        \end{tabular}
    }
    \endgroup
\end{table*}

Motivated by these critical gaps, this paper develops a unified analytical and algorithmic framework for OFDM-ISAC sensing beyond the CP limit, encompassing signal modeling, theoretical analysis, and interference mitigation algorithms. The main contributions are summarized as follows:

\begin{itemize}
    \item We establish a general OFDM-ISAC echo model that explicitly captures the structured coupling between ISI and ICI caused by an insufficient CP. This model reveals how CP insufficiency breaks the separable range-Doppler structure and introduces coherent interference across subcarriers and OFDM symbols, motivating our theoretical performance analysis and algorithm design.

    \item We derive closed-form analytical expressions for the sensing signal-to-interference-plus-noise ratio (SINR) and the second-order characterization of the RDM without invoking independence or Gaussianity assumptions on the interference terms. Based on this second-order characterization, we further obtain a tractable approximation for the peak sidelobe level ratio (PSLR). The analysis shows that both SINR degradation and sidelobe elevation increase approximately linearly with the normalized excess delay beyond the CP, providing quantitative insight into the trade-off between CP length and sensing performance.

    \item We propose two frequency-domain interference cancellation algorithms, SIC-DFT and SIC-ESPRIT, to mitigate ISI and ICI due to insufficient CP. SIC-DFT leverages low-complexity DFT-based estimation, while SIC-ESPRIT exploits subspace-based parameter recovery to achieve super-resolution sensing accuracy. Both methods are fully compatible with standard OFDM signaling and require no waveform modification.
\end{itemize}

\textcolor{black}{Extensive simulations are further provided to validate the theoretical analysis and to evaluate the proposed algorithms against representative benchmarks. The results demonstrate that both SIC-DFT and SIC-ESPRIT consistently outperform existing benchmark methods, achieving more than a $4$~dB SINR improvement under the same CP-insufficient conditions. In particular, SIC-ESPRIT reduces both range and velocity root-mean-square-errors (RMSEs) by approximately one order of magnitude compared with competing schemes, approaching the performance of systems employing a sufficiently long CP.}

\emph{Notation}: Boldface lowercase and uppercase letters denote vectors and matrices, respectively. The operators $(\cdot)^*$, $(\cdot)^T$, and $(\cdot)^H$ denote conjugate, transpose, and Hermitian transpose. The symbols $\otimes$, $\odot$, and $\|\cdot\|_F$ represent Kronecker product, Hadamard product, and Frobenius norm. The notation $\mathbb{E}\{\cdot\}$ denotes statistical expectation, $\mathcal{CN}(\mu, \sigma^2)$ represents complex Gaussian distribution, $\mathrm{diag}\{\cdot\}$ constructs a diagonal matrix from its vector argument, $\mathrm{vec}(\cdot)$ denotes vectorization, and $\mathbf{I}_N$ is the $N \times N$ identity matrix. We use $\jmath = \sqrt{-1}$ for the imaginary unit and $\triangleq$ for definitions.

\section{Signal Model and Echo Signal Processing} \label{sec:system_model}
\subsection{Transmit Signal Model}

Consider an OFDM frame with $M$ symbols and $N$ subcarriers, for which the baseband transmit signal including the CP can be expressed as
\begin{equation}
    x(t) = \sum_{m=0}^{M-1} \sum_{n=0}^{N-1}
    \frac{s_{n,m}}{\sqrt{N}}
    e^{\jmath 2\pi n \Delta_f (t - mT_{\mathrm{s}} - T_{\mathrm{cp}})} g\Big(\frac{t - mT_{\mathrm{s}}}{T_{\mathrm{s}}}\Big),
\end{equation}
where $s_{n,m}$ denotes the data symbol on the $n$-th subcarrier of the $m$-th OFDM symbol, $\Delta_f$ denotes the subcarrier spacing, the total OFDM symbol duration including the CP is $T_{\mathrm{s}} = T + T_{\mathrm{cp}}$ with CP duration $T_{\mathrm{cp}}$ and ``useful'' OFDM symbol duration (i.e., IFFT duration) $T = 1/\Delta_f$, and $g(\xi)$ is a rectangular pulse shaping function, i.e., $g(\xi)=1$ for $0\le \xi<1$ and $0$ otherwise. The data symbols are assumed to be independent across subcarriers and OFDM symbols, with $\mathbb{E}\{|s_{n,m}|^2\}=1$ and $\mathbb{E}\{s_{n,m}\}=0$.
\textcolor{black}{The condition $\mathbb{E}\{s_{n,m}^2\}=0$ holds for the QPSK and square QAM constellations considered in the paper, and more generally for many rotationally symmetric constellations.}

\begin{figure}[!t]
    \centering
    \includegraphics[width = 3.3 in]{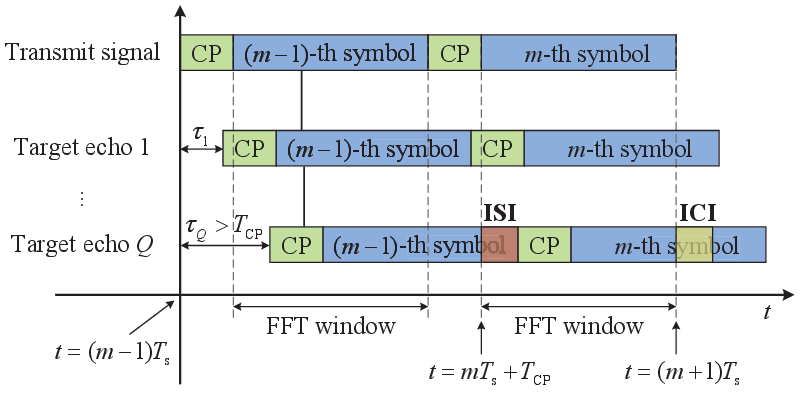}
    \caption{Illustration of transmit and echo signals. ISI and ICI occur when the CP duration $T_{\mathrm{cp}}$ is shorter than the maximum target delay $\tau_Q$.}    \label{fig:echo_signal}
    \vspace{-0.3 cm}
\end{figure}

\subsection{Radar Echo Signal Model}
The sensing receiver activates its detection window immediately after transmission to ensure that echoes from nearby targets are not missed. Suppose there exist $Q$ point targets at ranges $R_q$ with relative radial velocities $v_q$ and average radar cross sections (RCSs) $\sigma_{\mathrm{rcs}, q}$ for $q = 1, \dots, Q$. The echo signal at the sensing receiver is expressed as
\begin{equation} \label{eq:echo_sig}
    y(t) = \sum_{q=1}^{Q} \alpha_q x(t-\tau_q) e^{\jmath 2\pi f_{\mathrm{d}, q} t} + z(t),
\end{equation}
where $\alpha_q$ denotes the complex reflection coefficient of the $q$-th target and $z(t)$ is additive white Gaussian noise (AWGN). \textcolor{black}{Following the fluctuating-target model in \cite{Richards_book_2010,Keskin_TWC_2025}, we model the target reflection coefficients as independent circularly symmetric complex Gaussian random variables that remain constant over one OFDM frame, i.e., $\alpha_q \sim \mathcal{CN}(0,\sigma_{\alpha,q}^{2})$. The variance $\sigma_{\alpha,q}^{2}$, corresponding to the average received echo power, is determined by the monostatic radar equation as}
\begin{equation}
    \textcolor{black}{\sigma_{\alpha,q}^{2} = \frac{P_{\mathrm{t}} G_{\mathrm{t}}G_{\mathrm{r}}\lambda_{\mathrm{c}}^{2}\sigma_{\mathrm{rcs}, q}}{(4 \pi)^3 R_q^4},}
\end{equation}
\textcolor{black}{where $\lambda_{\mathrm{c}}=c_0/f_{\mathrm{c}}$ is the carrier wavelength, $P_{\mathrm{t}}$ is the transmit power, and $G_{\mathrm{t}}$ and $G_{\mathrm{r}}$ are the transmit and receive antenna gains, respectively.} The target delay and Doppler shift are given by
\begin{equation}
    \tau_q = \frac{2 R_q}{c_0}, ~~ f_{\mathrm{d}, q} = \frac{2 v_q f_{\mathrm{c}}}{c_0},
\end{equation}
where $c_0$ is the speed of light.

Following \cite{Berger_JSTSP_2010,Keskin_TWC_2025}, we neglect the intra-symbol Doppler-induced phase variation, while retaining the inter-symbol Doppler phase progression for velocity estimation. This approximation is accurate when the normalized Doppler shift $f_\text{d}/\Delta_f$ is small, as in typical 5G NR configurations \cite{3gpp_standards}. For instance, at a representative FR2/mmWave carrier frequency $f_{\mathrm{c}} = 28$~GHz, a target with velocity $v = 30$~m/s induces a Doppler shift of $f_{\mathrm{d}} = 5.6$~kHz, corresponding to $f_\text{d}/\Delta_f\approx 0.047$ for $\Delta_f=120$ kHz.
\textcolor{black}{For highly mobile targets such as high-speed UAVs, the normalized Doppler shift may become non-negligible and induce additional Doppler-related ICI. In such scenarios, the proposed CP-induced ISI/ICI model can be extended or combined with ICI-aware OFDM radar processing methods \cite{Hakobyan_TVT_2018, Keskin_JSTSP_2021}. This paper focuses on the complementary problem of CP-induced ISI/ICI.} Under this assumption, the echo signal in \eqref{eq:echo_sig} can be rewritten as
\begin{equation}
    \begin{aligned}
        y(t) & \approx \sum_{q=1}^{Q} \frac{\alpha_q}{\sqrt{N}}  \sum_{m=0}^{M-1} \sum_{n=0}^{N-1} s_{n,m} e^{\jmath 2\pi n \Delta_f (t - m T_{\mathrm{s}} - \tau_q - T_{\mathrm{cp}})} \\
             & ~~~ \times e^{\jmath 2\pi m f_{\mathrm{d},q} T_{\mathrm{s}} } g\Big(\frac{t - m T_{\mathrm{s}} - \tau_q}{T_{\mathrm{s}}}\Big) + z(t),
    \end{aligned}
\end{equation}

At the sensing receiver, the radar echo signal $y(t)$ is sampled every $1/B~\mathrm{s}$, where $B = N \Delta_f$ is the signal bandwidth. The resulting discrete-time echo signal is thus represented as
\begin{subequations}
    \begin{align}
        y [i] & = \sum_{q=1}^{Q} \frac{\alpha_q}{\sqrt{N}}  \sum_{m=0}^{M-1} \sum_{n=0}^{N-1} s_{n,m} e^{\jmath \frac{2\pi}{N} n (i - mN_{\mathrm{s}} - N_{\mathrm{cp}})}                         \\
              & ~~~ \times e^{-\jmath 2\pi n \Delta_f \tau_q}  e^{\jmath 2\pi m f_{\mathrm{d},q} T_{\mathrm{s}} } g\Big(\frac{i \!-\! mN_{\mathrm{s}}\!-\! l_q }{N_{\mathrm{s}}}\Big) \!+\! z[i],
    \end{align}
\end{subequations}
where $l_q = [\tau_q B]$, $N_{\mathrm{cp}} = T_{\mathrm{cp}}B$ denotes the CP length, $N_{\mathrm{s}} = N + N_{\mathrm{cp}}$, and $[\cdot]$ is the rounding operation. For the echo signal corresponding to the $q$-th target, the $m$-th symbol spans the sample indices from $i = m N_{\mathrm{s}} + l_q$ to $i = (m+1) N_{\mathrm{s}} + l_q -1$.
After CP removal, OFDM demodulation is applied to the $N$ samples from $i = m N_{\mathrm{s}} + N_{\mathrm{cp}}$ to $i = (m+1) N_{\mathrm{s}}-1$.

\begin{figure*}[!t]
    \begin{equation}
        \tilde{y}_{m, q} [i] = \! \frac{\alpha_q}{\sqrt{N}} \!\sum_{n=0}^{N-1}\! s_{n,m-1} e^{\jmath \frac{2\pi}{N} ni} e^{-\jmath 2\pi n \Delta_f (\tau_q - T_{\mathrm{cp}})}  e^{\jmath 2\pi (m-1) f_{\mathrm{d},q} T_{\mathrm{s}} } g_1(i) + \frac{\alpha_q}{\sqrt{N}} \! \sum_{n=0}^{N-1} \! s_{n,m} e^{\jmath \frac{2\pi}{N} n i} e^{-\jmath 2\pi n \Delta_f \tau_q}  e^{\jmath 2\pi m f_{\mathrm{d},q} T_{\mathrm{s}} } g_2(i).\!\label{eq:echo_time}
    \end{equation}
    \rule[-0pt]{18.1cm}{0.05em}
\end{figure*}

\subsubsection{Sufficient CP Length Case}
When the delay of the $q$-th target does not exceed the CP length, i.e., $l_q \leq N_{\mathrm{cp}}$, neither ISI nor ICI occurs. In this case, the detection window for the $m$-th symbol fully covers the entire duration of the current transmitted symbol (as exemplified by the echo of the first target in Fig.~\ref{fig:echo_signal}), ensuring subcarrier orthogonality and interference-free reception. The discrete-time echo signal for the $q$-th target can be expressed as
\begin{equation}
    \!\!y_{m,q} [i] = \!\sum_{n=0}^{N-1}\! \frac{\alpha_q s_{n,m}}{\sqrt{N}}  e^{\jmath \frac{2\pi}{N} n i}  e^{-\jmath 2\pi n \Delta_f \tau_q}  e^{\jmath 2\pi m f_{\mathrm{d},q} T_{\mathrm{s}} } g\Big(\!\frac{i}{N}\!\Big). \!\!\!
\end{equation}
After OFDM demodulation, the frequency-domain echo on the $n$-th subcarrier of the $m$-th symbol from the $q$-th target is
\begin{subequations}
    \begin{align}
        y_{n,m,q} & = \frac{1}{\sqrt{N}}\sum_{i=0}^{N-1} y_{m,q} [i] e^{-\jmath \frac{2\pi}{N} n i}                                                                                                                                \\
                  & = \!\frac{\alpha_q}{N} e^{\jmath 2\pi m f_{\mathrm{d},q} T_{\mathrm{s}} }  \!\! \sum_{n'=0}^{N-1} \!\! s_{n',m} e^{-\jmath 2\pi n' \Delta_f \tau_q} \! \sum_{i=0}^{N-1} \!e^{\jmath \frac{2\pi}{N} (n' - n) i} \\
                  & = \alpha_q s_{n,m} e^{-\jmath 2\pi n \Delta_f \tau_q} e^{\jmath 2\pi m f_{\mathrm{d},q} T_{\mathrm{s}} }.
    \end{align}
\end{subequations}

\subsubsection{Insufficient CP Length Case}
When the delay of the $q$-th target exceeds the CP length, i.e., $l_q>N_{\mathrm{cp}}$, both ISI and ICI arise and degrade sensing performance. \textcolor{black}{In this paper, we focus on the practical one-symbol-overlap case $N_{\mathrm{cp}}<l_q<N_{\mathrm{s}}$, where each beyond-CP echo overlaps with at most one adjacent OFDM symbol. Longer-delay echoes spanning multiple OFDM symbols can be handled by introducing higher-order temporal-shift matrices, although this extension is omitted for simplicity.} As illustrated in Fig.~\ref{fig:echo_signal}, for the echo of the $Q$-th target, the FFT window of the $m$-th symbol overlaps the tail of symbol $m-1$ (yielding ISI) and truncates the body of symbol $m$ (breaking subcarrier orthogonality and inducing ICI). In this case, the discrete-time echo of the $q$-th target within the $m$-th symbol is given by (\ref{eq:echo_time}) at the top of this page, where
$g_1(i)=g\big(\frac{i}{\,l_q-N_{\mathrm{cp}}\,}\big)$ selects the first $l_q-N_{\mathrm{cp}}$ samples (residual from symbol $m-1$), and $g_2(i)=g\big(\frac{i-l_q+N_{\mathrm{cp}}}{\,N-l_q+N_{\mathrm{cp}}\,}\big)$ selects the remaining $N-(l_q\!-\!N_{\mathrm{cp}})$ samples of symbol $m$. After OFDM demodulation, the corresponding frequency-domain echo $\tilde{y}_{n,m,q}$ can be expressed as\footnote{The negative sign before the ICI term originates from the time-domain signal construction: The actual received signal of the $m$-th symbol can be viewed as the interference-free signal minus the portion of samples that fall outside the CP.}
\begin{subequations}
    \begin{align}
        \!\!\!\!\!\!\tilde{y}_{n,m,q}
         & = \! \frac{1}{\sqrt{N}} \sum_{i=0}^{N-1} \tilde{y}_{m,q}[i] e^{-\jmath \frac{2\pi}{N} ni}                                                                                         \\
         & = \! \alpha_q s_{n,m} e^{-\jmath 2\pi n \Delta_f \tau_q} e^{\jmath 2\pi m f_{\mathrm{d},q} T_{\mathrm{s}} } \!+\! y_{n,m,q}^{\mathrm{ISI}} \!-\! y_{n,m,q}^{\mathrm{ICI}}, \!\!\!
    \end{align}
\end{subequations}
where $y^{\mathrm{ISI}}_{n,m,q}$ and $y^{\mathrm{ICI}}_{n,m,q}$ denote the ISI and ICI components, respectively:
\begin{subequations} \label{eq:ISI_ICI_expr}
    \begin{align}
        y_{n,m,q}^{\mathrm{ISI}} & \!=\! \alpha_q e^{\jmath 2\pi (m\!-\!1) f_{\mathrm{d},q} T_{\mathrm{s}} } \!\sum_{n'=0}^{N-1}\!\! s_{n',m\!-\!1} e^{\jmath 2\pi n' \Delta_f (T_{\mathrm{cp}} - \tau_q)} \phi^q_{n,n'}, \\
        y_{n,m,q}^{\mathrm{ICI}} & = \alpha_q e^{\jmath 2\pi m f_{\mathrm{d},q} T_{\mathrm{s}} }  \sum_{n'=0}^{N-1} s_{n',m} e^{-\jmath 2\pi n' \Delta_f \tau_q} \phi^q_{n,n'},
    \end{align}
\end{subequations}
where $\phi^q_{n,n'} = \frac{1}{N} \sum_{i=0}^{l_q-N_{\mathrm{cp}}-1} e^{\jmath \frac{2\pi}{N}(n'-n)i}$.

To distinguish between the interference-free and ISI/ICI-contaminated echoes, the targets are partitioned according to their round-trip delays. Without loss of generality, we order the targets such that $\tau_1 \le \cdots \le \tau_Q$ and assume that the first $\tilde{Q}$ targets have round-trip delays satisfying $l_q \le N_{\mathrm{cp}}$, for $q=1,\ldots,\tilde{Q}$. Thus, the frequency-domain echo on the $n$-th subcarrier of the $m$-th OFDM symbol is given by
\begin{subequations} \label{eq:sig_freq}
    \begin{align}
        y_{n,m} & = \sum_{q=1}^{\tilde{Q}} y_{n,m,q} + \sum_{q=\tilde{Q}+1}^Q \tilde{y}_{n,m,q}                                                                                                                                                                                      \\
                & = \underbrace{ \sum_{q=1}^Q  \alpha_q s_{n,m} e^{-\jmath 2\pi n \Delta_f \tau_q} e^{\jmath 2\pi m f_{\mathrm{d},q} T_{\mathrm{s}} } }_{\textnormal{ISI/ICI-free component}, ~y_{n,m}^{\mathrm{free}}} \notag                                                       \\
                & \hspace{2mm}+ \underbrace{ \sum_{q=\tilde{Q}+1}^Q y_{n,m,q}^{\mathrm{ISI}} }_{\textnormal{ISI component}, ~y_{n,m}^{\mathrm{ISI}}} - \underbrace{\sum_{q=\tilde{Q}+1}^Q y_{n,m,q}^{\mathrm{ICI}}}_{\textnormal{ICI component}, ~y_{n,m}^{\mathrm{ICI}}} + z_{n,m},
    \end{align}
\end{subequations}
where $z_{n,m} \sim \mathcal{CN}(0, \sigma^2)$ denotes AWGN, $\sigma^2 = F k_b \Delta_f T_{\mathrm{temp}}$ is the noise power, $F$ is the receiver's noise figure, $k_b$ Boltzmann's constant, and $T_{\mathrm{temp}}$ the equivalent noise temperature.

For notational compactness, we define the delay-domain steering vector, Doppler-domain steering vector, and ISI/ICI-induced phase matrix as
\begin{subequations}
    \begin{align}
        \mathbf{b}(\tau_q)           & \triangleq [1, e^{-\jmath 2\pi \Delta_f \tau_q}, \dots, e^{-\jmath 2\pi (N-1) \Delta_f \tau_q}]^T,                                   \\
        \mathbf{c}(f_{\mathrm{d},q}) & \triangleq [1, e^{-\jmath 2\pi f_{\mathrm{d},q} T_{\mathrm{s}}}, \dots,  e^{-\jmath 2\pi (M-1) f_{\mathrm{d},q} T_{\mathrm{s}}} ]^T, \\
        [\mathbf{\Phi}_q]_{n,n'}     & \triangleq \phi^q_{n,n'}.
    \end{align}
\end{subequations}
The ISI/ICI-free, ISI, and ICI components for the $m$-th symbol can thus be written compactly as
\begin{subequations} \label{eq:sig_vec}
    \begin{align}
        \!\!\!\mathbf{y}_m^{\mathrm{free}} & = \sum_{q=1}^Q \alpha_q \mathbf{b}(\tau_q) [\mathbf{c}^{\ast}(f_{\mathrm{d},q})]_m \odot \mathbf{s}_m,                                                                             \\
        \mathbf{y}_{m}^{\mathrm{ISI}}      & =\! \sum_{q=\tilde{Q}+1}^Q \!\!\!\alpha_q \mathbf{\Phi}_q \big(\mathbf{b}(\tau_q \!-\! T_{\mathrm{cp}})[\mathbf{c}^{\ast}(f_{\mathrm{d},q})]_{m-1} \odot \mathbf{s}_{m-1}\big), \! \\
        \mathbf{y}_m^{\mathrm{ICI}}        & =  \sum_{q=\tilde{Q}+1}^Q \alpha_q \mathbf{\Phi}_q \big(\mathbf{b}(\tau_q) [\mathbf{c}^{\ast}(f_{\mathrm{d},q})]_m \odot \mathbf{s}_m\big),
    \end{align}
\end{subequations}
where
\begin{subequations}
    \begin{align}
        \mathbf{y}_m^{\mathrm{free}} & \triangleq [y_{0,m}^{\mathrm{free}}, \dots, y_{N-1, m}^{\mathrm{free}}]^T, \\
        \mathbf{y}_m^{\mathrm{ISI}}  & \triangleq [y_{0,m}^{\mathrm{ISI}}, \dots, y_{N-1, m}^{\mathrm{ISI}}]^T,   \\
        \mathbf{y}_m^{\mathrm{ICI}}  & \triangleq [y_{0,m}^{\mathrm{ICI}}, \dots, y_{N-1, m}^{\mathrm{ICI}}]^T,   \\
        \mathbf{s}_m                 & \triangleq [s_{0,m}, \dots, s_{N-1,m}]^T.
    \end{align}
\end{subequations}
Aggregating (\ref{eq:sig_vec}) over $M$ symbols, the ISI/ICI-free, ISI, and ICI components over a frame can be further written in matrix form as
\begin{subequations} \label{eq:sig_matr}
    \begin{align}
        \mathbf{Y}_{\mathrm{free}} & = \sum_{q=1}^Q \alpha_q \big(\mathbf{b}(\tau_q)\mathbf{c}^H(f_{\mathrm{d},q}) \odot \mathbf{S} \big),                                                          \\
        \mathbf{Y}_{\mathrm{ISI}}  & = \sum_{q=\tilde{Q}+1}^Q \alpha_q \mathbf{\Phi}_q \big( \mathbf{b}(\tau_q-T_{\mathrm{cp}}) \mathbf{c}^H(f_{\mathrm{d},q}) \odot \mathbf{S} \big) \mathbf{J}_1, \\
        \mathbf{Y}_{\mathrm{ICI}}  & = \sum_{q=\tilde{Q}+1}^Q \alpha_q \mathbf{\Phi}_q \big( \mathbf{b}(\tau_q) \mathbf{c}^H(f_{\mathrm{d},q}) \odot \mathbf{S} \big),
    \end{align}
\end{subequations}
where $\mathbf{Y}_{\mathrm{free}} \triangleq [\mathbf{y}_{0}^{\mathrm{free}}, \dots, \mathbf{y}_{M-1}^{\mathrm{free}}]$, $\mathbf{Y}_{\mathrm{ISI}} \triangleq [\mathbf{y}_{0}^{\mathrm{ISI}}, \dots, \mathbf{y}_{M-1}^{\mathrm{ISI}}]$, $\mathbf{Y}_{\mathrm{ICI}} \triangleq [\mathbf{y}_{0}^{\mathrm{ICI}}, \dots, \mathbf{y}_{M-1}^{\mathrm{ICI}}]$, $ \mathbf{S} \triangleq [\mathbf{s}_{0}, \dots, \mathbf{s}_{M-1}]$, and $\mathbf{J}_1 \in \mathbb{C}^{M \times M}$ is a shift matrix defined as
\begin{equation}
    \mathbf{J}_1 \triangleq \begin{bmatrix}
        \mathbf{0}_{M-1} & \mathbf{I}_{M-1}   \\
        0                & \mathbf{0}^T_{M-1}
    \end{bmatrix}.
\end{equation}
Finally, the sensing observation matrix can be expressed as
\begin{equation}
    \mathbf{Y} = \mathbf{Y}_{\mathrm{free}} + \mathbf{Y}_{\mathrm{ISI}} - \mathbf{Y}_{\mathrm{ICI}} + \mathbf{Z},
\end{equation}
where the noise matrix $\mathbf{Z} \in \mathbb{C}^{N \times M}$ consists of independent and identically distributed (i.i.d.) complex Gaussian entries, i.e., $[\mathbf{Z}]_{n,m} = z_{n,m}$ with $\mathrm{vec}(\mathbf{Z}) \sim \mathcal{CN}(\mathbf{0}, \sigma^2 \mathbf{I}_{MN})$. Given the known data matrix $\mathbf{S}$, the objective of OFDM-based radar sensing is to jointly estimate the target parameters, i.e. the reflection coefficients $\{\alpha_q\}_{q=1}^Q$, delays $\{\tau_q\}_{q=1}^Q$, and Doppler shifts $\{f_{\mathrm{d},q}\}_{q=1}^Q$, from the observation matrix $\mathbf{Y}$.

\begin{figure}[!t]
    \centering
    \includegraphics[width = 2.8 in]{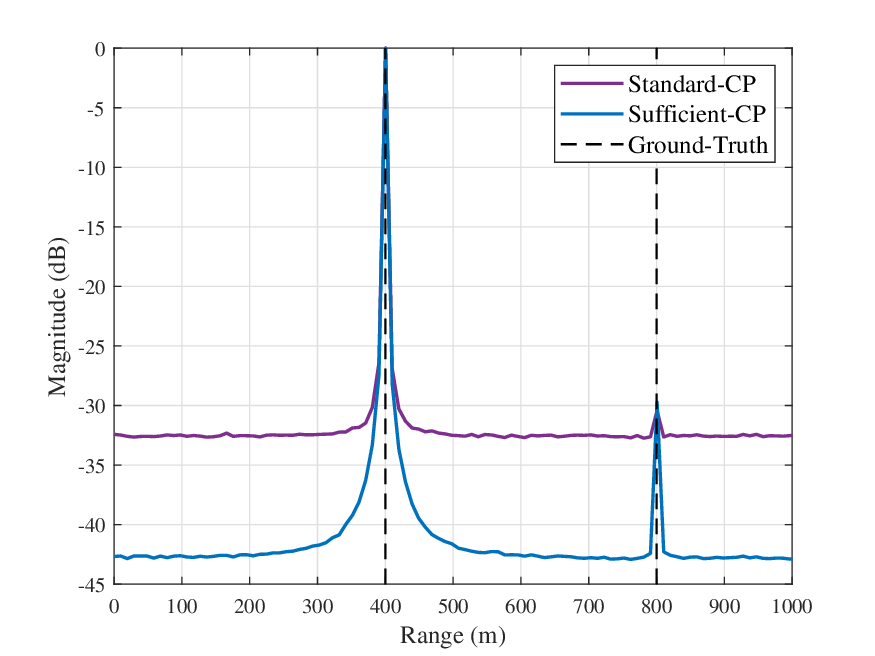}
    \caption{Range profiles generated using the 2D-DFT sensing algorithm under standard-CP and sufficient-CP configurations.} \label{fig:range_profile_tar}
    \vspace{-0.2 cm}
\end{figure}

\section{Sensing Performance Analysis} \label{sec:performance_analysis}
The signal model developed in Section \ref{sec:system_model} reveals that an insufficient CP fundamentally distorts the structure of the received echo signal, introducing ISI and ICI.
In this section, we analyze how CP insufficiency affects the sensing performance of OFDM-ISAC systems.
Specifically, we characterize the ISI/ICI-induced degradation in terms of the SINR and the sidelobe level of the RDM.
This analysis provides insight into the sensing impairment caused by an insufficient CP and establishes the theoretical foundation for the interference cancellation algorithms developed in Section~\ref{sec:SIC algorithm}.

\subsection{Impact of Insufficient CP}
The interference-free component $\mathbf{Y}_{\mathrm{free}}$ of the signal model in~(\ref{eq:sig_matr}) exhibits a separable structure, represented by a rank-$Q$ sum of outer products $\mathbf{b}(\tau_q)\mathbf{c}^H(f_{\mathrm{d},q})$ modulated by the data matrix $\mathbf{S}$. After matched filtering and 2D-DFT processing, this component produces a well-localized RDM, allowing clear resolution of the target delays and Doppler shifts.

In contrast, the interference components $\mathbf{Y}_{\mathrm{ISI}}$ and $\mathbf{Y}_{\mathrm{ICI}}$ depart from this desirable separable structure.
Both involve the phase-rotation matrix $\mathbf{\Phi}_q$, which induces off-diagonal spectral leakage across subcarriers and destroys the subcarrier orthogonality.
\textcolor{black}{Moreover, the temporal-shift matrix $\mathbf{J}_1$ in $\mathbf{Y}_{\mathrm{ISI}}$ couples adjacent OFDM symbols and thereby introduces ISI.} Consequently, the interference terms no longer follow the canonical delay-Doppler steering-vector outer-product structure of $\mathbf{Y}_{\mathrm{free}}$.
After matched filtering and 2D-DFT processing, these noncanonical components manifest as subcarrier leakage, degraded SINR, and elevated RDM sidelobe floors.

This phenomenon is illustrated in Fig.~\ref{fig:range_profile_tar}.
In this example\footnote{The simulation settings for this example are consistent with those in Table~\ref{Tab:parameters} in the simulation section.}, two targets are positioned at ranges of $400$~m and $800$~m. Under the \emph{standard-CP} configuration defined by 3GPP \cite{3gpp_standards}, the CP duration is only $0.59~\mu$s, corresponding to a CP-protected range of approximately $90$~m. Consequently, both target echoes arrive well beyond the CP guard interval, inevitably causing ISI and ICI. For reference, we introduce a customized \emph{sufficient-CP} configuration, whose CP duration is extended to $8.33~\mu$s, equal to one OFDM symbol period, resulting in a CP-protected range of approximately $1250$~m. Under this sufficient-CP scenario, all target echoes arrive within the CP duration, entirely eliminating ISI/ICI and thus providing an ideal performance baseline. Fig.~\ref{fig:range_profile_tar} clearly demonstrates that the standard-CP scenario produces significantly elevated sidelobe levels compared to the sufficient-CP baseline, highlighting the substantial performance degradation induced by CP insufficiency. To rigorously quantify these detrimental impacts, the following sections derive analytical expressions for the resulting SINR and sidelobe characteristics, laying the foundation for our proposed interference-mitigation algorithms.

\subsection{SINR Analysis}
The SINR of the echo signal $\mathbf{Y}$ is defined as the ratio between the interference-free signal power and the combined interference and noise powers:
\begin{equation} \label{eq:SINR_defi}
    \mathrm{SINR} \triangleq \frac{\mathbb{E}\{\|\mathbf{Y}_{\mathrm{free}}\|_{F}^2\}} {\mathbb{E}\{\|\mathbf{Y}_{\mathrm{ISI}} - \mathbf{Y}_{\mathrm{ICI}}\|_{F}^2\} + \mathbb{E}\{\|\mathbf{Z}\|_{F}^2\}}.
\end{equation}
This metric characterizes the relative strength of the desired echo component $\mathbf{Y}_{\mathrm{free}}$ compared with the interference caused by $\mathbf{Y}_{\mathrm{ISI}}$, $\mathbf{Y}_{\mathrm{ICI}}$, and $\mathbf{Z}$.
It serves as a key indicator of the degradation in delay and Doppler estimation accuracy when the CP is insufficient.
Based on the signal model in (\ref{eq:sig_matr}), the SINR can be derived in closed form as follows.
\begin{prop} \label{prop:SINR}
    \textcolor{black}{The SINR of the echo signal $\mathbf{Y}$ is given by}
    \begin{equation} \label{eq:SINR_closed}
        \textcolor{black}{\mathrm{SINR} = \frac{M N\sum_{q=1}^Q \sigma_{\alpha,q}^{2}}{(2M-1)N \sum_{q=\tilde{Q}+1}^Q \rho_q \sigma_{\alpha,q}^{2} + M N\sigma^2},}
    \end{equation}
    where $\rho_q = (l_q - N_{\mathrm{cp}}) / N$ is the normalized excess delay beyond the CP for the $q$-th target.
\end{prop}
\noindent \emph{Proof:} See Appendix \ref{app:SINR_proof}.
\hfill $\blacksquare$
\smallskip

\textcolor{black}{The above expression is derived without assuming the statistical independence of ISI/ICI or approximating them as Gaussian noise, and thus provides a general characterization of SINR performance under CP insufficiency.}
To gain further insight, note that for a large number of OFDM symbols ($M \gg 1$), the SINR in (\ref{eq:SINR_closed}) is approximately
\begin{equation} \label{eq:SINR_asymp}
    \mathrm{SINR} \approx \frac{\sum_{q=1}^Q \sigma_{\alpha,q}^{2}}{2\sum_{q=\tilde{Q}+1}^Q \rho_q \sigma_{\alpha,q}^{2} + \sigma^2}.
\end{equation}
This expression highlights the pivotal role of the normalized excess delay beyond the CP, $\rho_q = (l_q - N_{\mathrm{cp}})/N$, in determining SINR performance.
If all echoes lie within the CP, the interference term vanishes and the SINR reduces to the noise-limited SNR. For fixed target reflection powers, the CP-induced interference contribution of a beyond-CP target is proportional to its normalized excess delay $\rho_q>0$. Therefore, as $\rho_q$ increases, the SINR decreases monotonically under the considered model.

\subsection{Sidelobe Level Analysis}
While SINR quantifies the overall performance degradation, it does not capture the spatial distribution of interference in the delay-Doppler domain. We now analyze how CP insufficiency elevates sidelobe levels in the RDM, which directly impacts the ability to detect weak targets. The RDM obtained using the 2D-DFT can be expressed as \cite{Pucci_JSAC_2022, Xiao_TSP_2024}
\begin{equation}\label{eq:RDM}
    \boldsymbol{\chi} = \mathbf{F}_N^H( \mathbf{Y} \odot \mathbf{S}^{\ast}) \mathbf{F}_M,
\end{equation}
where $\mathbf{F}_M$ and $\mathbf{F}_N^H$ denote the normalized DFT and inverse-DFT (IDFT) matrices, respectively. Since $\boldsymbol{\chi}$ is random due to data modulation, target reflection coefficients, and noise, the sensing performance is evaluated using its second-order moment $\mathbb{E}\{|\chi(l,\nu)|^2\}$.

\begin{prop} \label{prop:RDM}
    \textcolor{black}{The second-order moment of the RDM is
        \begin{equation}
            \mathbb{E}\big\{ |\chi(l,\nu)|^2 \big\} = \sum_{q = 1}^Q \frac{\tilde{\sigma}_{\alpha,q}^{2}}{MN} |D_N(l \!-\! \tilde{l}_q)|^2 |D_M(\nu \!-\! \tilde{\nu}_q)|^2 + \sigma_{\mathrm{SL}}^2,
        \end{equation}}
    where $D_N(x) = \frac{\sin(\pi x)}{\sin(\pi x/N)} e^{\jmath \pi (N-1)x/N}$ denotes the Dirichlet kernel, $\tilde{l}_q = \tau_q B$ and $\tilde{\nu}_q = f_{\mathrm{d},q} M T_{\mathrm{s}}$ represent the normalized delay and Doppler, $\mu_4 = \mathbb{E}\{|s_{n,m}|^4\}$, and \textcolor{black}{the effective reflection-coefficient variance $\tilde{\sigma}_{\alpha,q}^{2}$ is defined as}
    \begin{equation}
        \textcolor{black}{\tilde{\sigma}_{\alpha,q}^{2} \triangleq \begin{cases}
                \sigma_{\alpha,q}^{2},             & ~~q = 1, \dots, \widetilde{Q};   \\
                (1-\rho_q)^2\sigma_{\alpha,q}^{2}, & ~~q = \widetilde{Q}+1, \dots, Q.
            \end{cases}
        }
    \end{equation}
    \textcolor{black}{In addition, the constant sidelobe-floor term is given by
        \begin{equation} \label{eq:sigma_IN}
            \begin{aligned}
                \sigma_{\mathrm{SL}}^2
                = & (\mu_4 - 1) \sum_{q=1}^{\tilde{Q}} \sigma_{\alpha,q}^{2} + \sum_{q=\tilde{Q}+1}^Q \big((\mu_4 - 1)(1-\rho_q)^2                       \\
                  & \hspace{2.5cm} + \rho_q (2-\rho_q)\big) \sigma_{\alpha,q}^{2} + \sigma^2                                                             \\
                = & (\mu_4 - 1)\sum_{q=1}^{Q}\tilde{\sigma}_{\alpha,q}^{2} + \sum_{q=\tilde{Q}+1}^{Q}\rho_q(2-\rho_q) \sigma_{\alpha,q}^{2} + \sigma^2 ,
            \end{aligned}
        \end{equation}}
\end{prop}
\noindent \emph{Proof:} See Appendix \ref{app:RDM_proof}.
\hfill $\blacksquare$
\smallskip

\textcolor{black}{The sidelobe floor in (\ref{eq:sigma_IN}) consists of three terms. The first term is induced by modulation randomness and it vanishes for constant-modulus symbols such as QPSK. For a beyond-CP target, the fixed FFT window captures only a fraction $(1-\rho_q)$ of the coherent echo amplitude, leading to the reduced effective variance $\tilde{\sigma}_{\alpha,q}^{2}=(1-\rho_q)^2\sigma_{\alpha,q}^{2}$. The second term is the residual CP-induced ISI/ICI leakage after the coherent component has been accounted for. This term depends on the normalized excess delay $\rho_q$ and target power $\sigma_{\alpha,q}^{2}$, but not on $\mu_4$ under the unit-power independent-symbol assumptions. The third term is the noise floor. Therefore, although the residual ISI/ICI leakage term itself is modulation-independent in the second-order average, the overall sidelobe floor under CP insufficiency still depends on the modulation statistics through the first term.}

When all target normalized delay-Doppler pairs $(\tilde{l}_q, \tilde{\nu}_q)$ lie on distinct integer DFT bins, $\mathbb{E}\{ |\chi(l,\nu)|^2 \}$ simplifies to
\begin{equation} \label{eq:MF_output}
    \!\mathbb{E}\{|\chi(l, \nu)|^2\} = \begin{cases}
        MN \tilde{\sigma}_{\alpha,q}^{2} + \sigma_{\mathrm{SL}}^2, & (l, \nu) = (\tilde{l}_q, \tilde{\nu}_q); \\
        \sigma_{\mathrm{SL}}^2 ,                                   & \mathrm{otherwise}.
    \end{cases}
\end{equation}
\textcolor{black}{To further evaluate the sidelobe level, we define the PSLR for the $q$-th target as the normalized peak sidelobe level \cite{Song_TSP_2016, Alaie_SP_2019}:
\begin{equation}
    \gamma_q \triangleq \frac{\mathbb{E}\big\{\max_{(l,\nu)\in \mathcal{R}_{\mathrm{s}}} |\chi(l,\nu)|^2 \big\}}{\mathbb{E}\{ |\chi(\tilde{l}_q,\tilde{\nu}_q)|^2 \}},
\end{equation}
where $\mathcal{R}_{\mathrm{s}} = \{ (l, \nu) \mid (l, \nu) \neq (\tilde{l}_q, \tilde{\nu}_q), \forall q \}$ denotes the sidelobe region excluding the true target bins. Under this definition, $\gamma_q$ is a sidelobe-to-mainlobe power ratio; hence a smaller value indicates better sidelobe suppression.}

\textcolor{black}{For the PSLR approximation, we adopt the on-grid assumption that the target locations $(\tilde{l}_q, \tilde{\nu}_q)$ are mutually distinct integer DFT bins, and we evaluate the RDM over the $N\times M$ DFT grid. Since each sidelobe cell is formed by summing many independently modulated subcarrier-symbol contributions with small individual weights, the central limit theorem motivates the marginal Gaussian approximation $\chi(l,\nu)\sim\mathcal{CN}(0,\sigma_{\mathrm{SL}}^2)$ for each $(l,\nu)\in\mathcal{R}_{\mathrm{s}}$ when $MN$ is sufficiently large \cite{Papoulis_book_2002}.
For tractability, the sidelobe samples are further approximated as independent after the marginal Gaussian approximation, so that their squared magnitudes follow exponential distributions with mean $\sigma_{\mathrm{SL}}^2$.
According to \cite{Arnold_book_1992, Eisenberg_Statist_2008}, the expected peak sidelobe level can be approximated as
\begin{equation}
    \mathbb{E}\big\{\max_{(l,\nu)\in \mathcal{R}_{\mathrm{s}}} |\chi(l,\nu)|^2 \big\} \approx H_{|\mathcal{R}_{\mathrm{s}}|}\sigma_{\mathrm{SL}}^2,
\end{equation}
where $H_{|\mathcal{R}_{\mathrm{s}}|} = \sum_{r=1}^{|\mathcal{R}_{\mathrm{s}}|}1/r$ denotes the harmonic number and $|\mathcal{R}_{\mathrm{s}}|=MN-Q$ for integer delay-Doppler target bins.
The resulting PSLR is therefore approximated by
\begin{equation} \label{eq:PSLR_final}
    \gamma_q \approx \frac{H_{|\mathcal{R}_{\mathrm{s}}|}\sigma_{\mathrm{SL}}^2}{MN \tilde{\sigma}_{\alpha,q}^{2} + \sigma_{\mathrm{SL}}^2}.
\end{equation}}

From (\ref{eq:PSLR_final}), it is evident that an insufficient CP, characterized by the normalized excess delay $\rho_q$, concurrently raises the sidelobe floor and weakens the mainlobe return. As $\rho_q$ increases (i.e., a larger fraction of the echo falls outside the CP guard interval), the leakage of echo energy into the interference increases the noise floor $\sigma^2_{\mathrm{SL}}$ (boosting the sidelobe baseline), while the effective target variance is reduced to $\tilde{\sigma}_{\alpha,q}^{2}=(1-\rho_q)^2\sigma_{\alpha,q}^{2}$, diminishing the captured mainlobe power. Thus the PSLR $\gamma_q$ in (\ref{eq:PSLR_final}) increases monotonically with CP insufficiency.

\section{Iterative Algorithms for Beyond-CP Sensing}
\label{sec:SIC algorithm}
Building on the analytical results in Section \ref{sec:performance_analysis}, here we develop practical sensing algorithms that mitigate the ISI and ICI induced by an insufficient CP. The proposed framework adopts an SIC strategy that iteratively estimates target parameters, reconstructs the interference, and refines the interference-free echo signal. Two implementations are presented: a low-complexity SIC-DFT algorithm based on conventional 2D-DFT processing, and a high-resolution SIC-ESPRIT algorithm that exploits subspace-based parameter estimation.

\subsection{SIC-DFT: Efficient Iterative Cancellation}
For SIC-DFT, the interference-free observation is initialized as $\mathbf{Y}_{\mathrm{free}}^{0} = \mathbf{Y}$.
At the $k$-th iteration, the RDM is obtained using
\begin{equation} \label{eq:2D-DFT}
    \boldsymbol{\chi}^{k}
    = \mathbf{F}_N^{H} (\mathbf{Y}_{\mathrm{free}}^{k} \odot \mathbf{S}^{\ast}) \mathbf{F}_M .
\end{equation}
A constant false alarm rate (CFAR) detector is then applied to $\boldsymbol{\chi}^{k}$ to identify the targets and obtain the corresponding delay and Doppler estimates $\hat{\tau}_q^{k}$ and $\hat{f}_{\mathrm{d},q}^{k}$.
Given these estimates, the reflection coefficient of the $q$-th target is estimated via least squares (LS) \cite{Keskin_radarconf_2025}:
\begin{equation} \label{eq:alpha_esti}
    \hat{\alpha}_q^{k} = \frac{\mathbf{b}^{H}(\hat{\tau}_q^{k})  (\mathbf{Y}_{\mathrm{free}}^{k} \odot \mathbf{S}^{\ast}) \mathbf{c}(\hat{f}_{\mathrm{d},q}^{k})}{\| \mathbf{S} \|_{F}^{2}} .
\end{equation}

Using the current parameter estimates, the ISI and ICI components are reconstructed as
\begin{subequations} \label{eq:ISI_ICI_recon}
    \begin{align}
        \mathbf{Y}_{\mathrm{ISI}}^k & = \sum_{q=\tilde{Q}+1}^Q \hat{\alpha}_q^k \mathbf{\Phi}_q \big( \mathbf{b}(\hat{\tau}_q^k-T_{\mathrm{cp}}) \mathbf{c}^H(\hat{f}_{\mathrm{d},q}^k) \odot \mathbf{S} \big) \mathbf{J}_1, \\
        \mathbf{Y}_{\mathrm{ICI}}^k & = \sum_{q=\tilde{Q}+1}^Q \hat{\alpha}_q^k \mathbf{\Phi}_q \big( \mathbf{b}(\hat{\tau}_q^k) \mathbf{c}^H(\hat{f}_{\mathrm{d},q}^k) \odot \mathbf{S} \big),
    \end{align}
\end{subequations}
and the interference-free observation is then updated as
\begin{equation} \label{eq:sub_isi_ici}
    \mathbf{Y}_{\mathrm{free}}^{k+1} = \mathbf{Y} - \mathbf{Y}_{\mathrm{ISI}}^k + \mathbf{Y}_{\mathrm{ICI}}^k.
\end{equation}
This process repeats until convergence, determined by
\begin{equation} \label{eq:stop_rule}
    \frac{\|\mathbf{Y}_{\mathrm{free}}^{k+1} - \mathbf{Y}_{\mathrm{free}}^{k}\|_F}{\|\mathbf{Y}_{\mathrm{free}}^{k}\|_F} < \delta_{\mathrm{th}} ~\text{or}~ k \geq K_{\max},
\end{equation}
where $\delta_{\mathrm{th}}$ and $K_{\max}$ denote the convergence tolerance and maximum iteration count, respectively.

\begin{algorithm}[!t]
    \begin{small}
        \caption{SIC-DFT Algorithm}
        \label{Alg:SIC_DFT}
        \begin{algorithmic}[1]
            \REQUIRE {$\mathbf{Y}$, $\mathbf{S}$, $\delta_{\mathrm{th}}$, $K_{\max}$.}
            \ENSURE {$\hat{\tau}_q^{\star}$, $\hat{f}_{\mathrm{d},q}^{\star}$, $\forall q$}
            \STATE {Initialize the interference-free signal $\mathbf{Y}_{\mathrm{free}}^{0} = \mathbf{Y}$, $k=0$.}
            \REPEAT
            \STATE {Compute the RDM from $\mathbf{Y}_{\mathrm{free}}^{k}$ using (\ref{eq:2D-DFT}).}
            \STATE {Estimate delay $\hat{\tau}_q^{k}$ and Doppler $\hat{f}_{\mathrm{d},q}^{k}$ with a CFAR detector.}
            \STATE {Estimate target reflection coefficient $\hat{\alpha}_q^{k}$ using (\ref{eq:alpha_esti}).}
            \STATE {Reconstruct ISI and ICI components $\mathbf{Y}_{\mathrm{ISI}}^k$, $\mathbf{Y}_{\mathrm{ICI}}^k$ using (\ref{eq:ISI_ICI_recon})}.
            \STATE {Compute the interference-free signal $\mathbf{Y}_{\mathrm{free}}^{k+1}$ using (\ref{eq:sub_isi_ici}).}
            \STATE {Set $k:=k+1$.}
            \UNTIL {$\|\mathbf{Y}_{\mathrm{free}}^{k}-\mathbf{Y}_{\mathrm{free}}^{k-1}\|_F/\|\mathbf{Y}_{\mathrm{free}}^{k}\|_F<\delta_{\mathrm{th}}$ or $k \ge K_{\max}$}
            \STATE {Return $\hat{\tau}_q^{\star} = \hat{\tau}_q^{k-1}$, $\hat{f}_{\mathrm{d},q}^{\star} = \hat{f}_{\mathrm{d},q}^{k-1}$, $\forall q$.}
        \end{algorithmic}
    \end{small}
\end{algorithm}

The SIC-DFT procedure is summarized in Algorithm \ref{Alg:SIC_DFT}.
Each iteration requires a 2D-DFT and ISI/ICI reconstruction.
\textcolor{black}{With an FFT implementation, the 2D-DFT requires computation of order $\mathcal{O}\big(MN(\log N+\log M)\big)$, while reconstructing the ISI/ICI terms for $Q$ targets can be implemented by FFT-based multiplication along the subcarrier dimension with complexity $\mathcal{O}(QMN\log N)$. Hence, the overall complexity is $\mathcal{O}\big(N_{\mathrm{iter}}MN(Q\log N+\log M)\big)$, where $N_{\mathrm{iter}}$ is the number of iterations.}
This cost remains moderate for typical ISAC configurations, making SIC-DFT suitable for real-time implementation.

\begin{algorithm}[!t]
    \begin{small}
        \caption{SIC-ESPRIT Algorithm}
        \label{Alg:SIC_ESPRIT}
        \begin{algorithmic}[1]
            \REQUIRE {$\mathbf{Y}$, $\mathbf{S}$, $\delta_{\mathrm{th}}$, $K_{\max}$.}
            \ENSURE {$\hat{\tau}_q^{\star}$, $\hat{f}_{\mathrm{d},q}^{\star}$, $\forall q$}
            \STATE {Initialize the interference-free signal $\mathbf{Y}_{\mathrm{free}}^{0} = \mathbf{Y}$, $k=0$.}
            \REPEAT
            \STATE {Estimate the sensing channel by $\hat{\mathbf{h}} = \mathrm{vec}(\mathbf{Y}_{\mathrm{free}}^{k} \odot \mathbf{S}^{\ast})$.}
            \STATE {Apply spatial smoothing to generate multiple snapshots and construct the sample covariance matrix $\hat{\mathbf{R}}$.}
            \STATE {Estimate the model order $Q$ using MDL/AIC.}
            \STATE {Perform EVD of $\hat{\mathbf{R}}$ and obtain the signal subspace $\mathbf{U}_{\mathrm{s}}$.}
            \STATE {Compute $\mathbf{P}_{\tau}$ and $\mathbf{P}_f$ using (\ref{eq:P_tau}) and (\ref{eq:P_f}).}
            \STATE {Estimate paired $\hat{\tau}_q^{k}$ and $\hat{f}_{\mathrm{d},q}^{k}$ using (\ref{eq:delay_dopp_est}).}
            \STATE {Estimate target reflection coefficient $\hat{\alpha}_q^{k}$ using (\ref{eq:alpha_esti}).}
            \STATE {Reconstruct ISI and ICI components $\mathbf{Y}_{\mathrm{ISI}}^k$, $\mathbf{Y}_{\mathrm{ICI}}^k$ using (\ref{eq:ISI_ICI_recon})}.
            \STATE {Compute the interference-free signal $\mathbf{Y}_{\mathrm{free}}^{k+1}$ using (\ref{eq:sub_isi_ici}).}
            \STATE {Set $k:=k+1$.}
            \UNTIL {$\|\mathbf{Y}_{\mathrm{free}}^{k}-\mathbf{Y}_{\mathrm{free}}^{k-1}\|_F/\|\mathbf{Y}_{\mathrm{free}}^{k}\|_F<\delta_{\mathrm{th}}$ or $k \ge K_{\max}$}
            \STATE Return $\hat{\tau}_q^{\star} = \hat{\tau}_q^{k-1}$, $\hat{f}_{\mathrm{d},q}^{\star} = \hat{f}_{\mathrm{d},q}^{k-1}$, $\forall q$.
        \end{algorithmic}
    \end{small}
\end{algorithm}

\subsection{SIC-ESPRIT}
Although SIC-DFT can suppress CP-induced interference with low complexity, its accuracy is limited by the grid resolution and spectral leakage of DFT-based processing.
To overcome these limitations, the SIC-ESPRIT algorithm replaces the DFT stage with a high-resolution parameter estimation method based on ESPRIT. \textcolor{black}{While other high-resolution methods could be adopted here, we employ ESPRIT because it provides a closed-form solution and avoids the two-dimensional spectral peak search required by methods such as MUSIC, which substantially reduces the computational burden.} A key insight that enables ESPRIT to be applied despite the ISI/ICI corruption is that, although the interference terms alter the covariance structure, they preserve the fundamental subspace spanned by the target steering vectors. This preservation is formalized as follows:
\begin{prop}\label{prop:cov_matrix}
    The ensemble covariance matrix of the sensing channel estimate $\hat{\mathbf{h}} = \mathrm{vec}(\mathbf{Y}\odot\mathbf{S}^{\ast})$, accounting for the presence of ISI and ICI, is
    \begin{equation}
        \mathbf{R}=\mathbb{E}\{\hat{\mathbf{h}}\hat{\mathbf{h}}^{H}\}
        =\mathbf{A}_Q\boldsymbol{\Sigma}_{\alpha}\mathbf{A}_Q^{H}
        +\sigma_{\mathrm{SL}}^{2}\mathbf{I}_{MN}, \label{eq:cov_theo}
    \end{equation}
    where $\mathbf{A}_Q=[\mathbf{a}_1,\ldots,\mathbf{a}_Q]\!\in\!\mathbb{C}^{MN\times Q}$ with $\mathbf{a}_q=\mathbf{c}^{\ast}(f_{\mathrm{d},q})\!\otimes\!\mathbf{b}(\tau_q)$, $\boldsymbol{\Sigma}_{\alpha}=\mathrm{diag}\{\tilde{\sigma}_{\alpha,1}^{2},\dots,\tilde{\sigma}_{\alpha,Q}^{2}\}$.
\end{prop}
\noindent\emph{Proof:} See Appendix \ref{app:covMatr_proof}.
\hfill$\blacksquare$
\smallskip

Proposition~3 reveals that the ISI/ICI manifests itself as a power scaling (through $\tilde{\sigma}_{\alpha,q}^{2}$) and noise enhancement (through $\sigma_{\mathrm{SL}}^2$), but preserves the column space of $\mathbf{A}_Q$. Therefore, subspace methods remain applicable for parameter extraction.
Given the preserved subspace structure, we apply ESPRIT to extract high-resolution delay-Doppler estimates. The procedure leverages the shift-invariance properties of the steering vectors, which are briefly outlined below.

\subsubsection{Subspace Extraction} When the number of targets is unknown, the model order $Q$ can be estimated from the eigenvalue profile using standard information criteria such as minimum description length (MDL) or Akaike information criterion (AIC). Perform eigenvalue decomposition of the covariance matrix:
\begin{equation} \label{eq:cov_evd}
    \mathbf{R} = \mathbf{U}_{\mathrm{s}} \boldsymbol{\Lambda}_{\mathrm{s}} \mathbf{U}_{\mathrm{s}}^H + \sigma_{\mathrm{SL}}^2 \mathbf{U}_{\mathrm{n}} \mathbf{U}_{\mathrm{n}}^H,
\end{equation}
where $\mathbf{U}_{\mathrm{s}} \in \mathbb{C}^{MN \times Q}$ spans the signal subspace corresponding to the $Q$ largest eigenvalues.

\subsubsection{Shift-Invariance Exploitation} Define selection matrices that extract overlapping subarrays:
\begin{subequations} \label{eq:selection_matrices}
    \begin{align}
        \mathbf{J}_0      & = [\mathbf{I}_{M-1}, \mathbf{0}] \otimes [\mathbf{I}_{N-1}, \mathbf{0}], \\
        \mathbf{J}_{\tau} & = [\mathbf{I}_{M-1}, \mathbf{0}] \otimes [\mathbf{0}, \mathbf{I}_{N-1}], \\
        \mathbf{J}_{f}    & = [\mathbf{0}, \mathbf{I}_{M-1}] \otimes [\mathbf{I}_{N-1}, \mathbf{0}].
    \end{align}
\end{subequations}
These matrices create subspace pairs related by diagonal phase shifts that encode the target parameters.

\subsubsection{Parameter Recovery} Form the rotation matrices:
\begin{subequations}
    \begin{align}
        \mathbf{P}_{\tau} & = (\tilde{\mathbf{U}}_{\mathrm{s}}^H \tilde{\mathbf{U}}_{\mathrm{s}})^{-1}
        \tilde{\mathbf{U}}_{\mathrm{s}}^H \mathbf{U}_{\mathrm{s}}^{\tau}, \label{eq:P_tau}             \\
        \mathbf{P}_{f}    & = (\tilde{\mathbf{U}}_{\mathrm{s}}^H \tilde{\mathbf{U}}_{\mathrm{s}})^{-1}
        \tilde{\mathbf{U}}_{\mathrm{s}}^H \mathbf{U}_{\mathrm{s}}^{f}, \label{eq:P_f}
    \end{align}
\end{subequations}
where $\tilde{\mathbf{U}}_{\mathrm{s}} = \mathbf{J}_0 \mathbf{U}_{\mathrm{s}}$, $\mathbf{U}_{\mathrm{s}}^{\tau} = \mathbf{J}_{\tau} \mathbf{U}_{\mathrm{s}}$, and $\mathbf{U}_{\mathrm{s}}^{f} = \mathbf{J}_{f} \mathbf{U}_{\mathrm{s}}$.
For 2D parameter pairing, let
\begin{equation}
    \mathbf{P}_{\tau}=\mathbf{T}\boldsymbol{\Lambda}_{\tau}\mathbf{T}^{-1}, \quad \boldsymbol{\Lambda}_{f}=\mathbf{T}^{-1}\mathbf{P}_{f}\mathbf{T}.
\end{equation}
The $q$-th diagonal elements of $\boldsymbol{\Lambda}_{\tau}$ and $\boldsymbol{\Lambda}_{f}$ correspond to the same target because the two rotation matrices share the same signal-subspace basis. The paired eigenvalues then yield the parameter estimates:
\begin{equation}
    \hat{\tau}_q = -\frac{\angle \lambda_q^{\tau}}{2\pi \Delta_f}, \qquad \hat{f}_{\mathrm{d},q} = \frac{\angle \lambda_q^{f}}{2\pi T_{\mathrm{s}}}, \label{eq:delay_dopp_est}
\end{equation}
where $\lambda_q^{\tau}=[\boldsymbol{\Lambda}_{\tau}]_{q,q}$ and $\lambda_q^{f}=[\boldsymbol{\Lambda}_{f}]_{q,q}$ are the paired eigenvalues associated with the $q$-th target.

\begin{table}[!t]
    \begingroup
    \footnotesize
    \centering
    \caption{\textcolor{black}{Computational Complexity Comparison}} \label{Tab:complexity}
    \textcolor{black}{%
        \begin{tabular}{c c}
            \toprule
            \textbf{Method}                                         & \textbf{Dominant complexity}                               \\
            \midrule
            DFT                                                     & $\mathcal{O}\big(MN(\log N+\log M)\big)$                   \\
            ESPRIT                                                  & $\mathcal{O}\big(MN L^2\big)$                              \\
            TDCC\cite{Wang_WiOpt_2023}, FDCC \cite{Geiger_SCC_2025} & $\mathcal{O}\big(MN(\log N+\log M)\big)$                   \\
            MTCC \cite{Geiger_SCC_2025}                             & $\mathcal{O}\big(MN(\log N+\log M+Q)\big)$                 \\
            SIC-DFT                                                 & $\mathcal{O}\big(N_{\mathrm{iter}}MN(Q\log N+\log M)\big)$ \\
            SIC-ESPRIT                                              & $\mathcal{O}\big(N_{\mathrm{iter}}MN(L^2+Q\log N)\big)$    \\
            \bottomrule
        \end{tabular}
    }
    \vspace{-2mm}
    \endgroup
\end{table}

Building on the ESPRIT-based delay-Doppler estimation, the proposed SIC-ESPRIT algorithm adopts the same iterative reconstruction-cancellation framework as SIC-DFT.
Its main advantage lies in the super-resolution capability of ESPRIT, which enables accurate estimation of closely spaced targets and effectively mitigates spectral leakage, resulting in more precise ISI/ICI reconstruction and improved sensing performance.
In practice, the theoretical covariance matrix $\mathbf{R}$ is estimated from multiple snapshots obtained via spatial smoothing, which enhances robustness by exploiting the shift-invariance property of the steering vectors \cite{Berger_JSTSP_2010}.
The complete procedure is summarized in Algorithm \ref{Alg:SIC_ESPRIT}.

From a computational perspective, the dominant cost of SIC-ESPRIT arises from subspace extraction, efficiently implemented using a singular value decomposition of the spatially smoothed data matrix.
This step has a complexity of $\mathcal{O}(MN L^2)$, where $L$ denotes the number of spatial-smoothing snapshots. Including ISI/ICI reconstruction, SIC-ESPRIT has overall complexity $\mathcal{O}\big(N_{\mathrm{iter}}MN(L^2+Q\log N)\big)$.

\textcolor{black}{Table~\ref{Tab:complexity} summarizes the dominant computational costs of the algorithms considered in the simulations. Compared with conventional FFT-based processing and coherent-compensation benchmarks, SIC-DFT only introduces a moderate iterative reconstruction overhead. SIC-ESPRIT deliberately incurs a higher subspace-processing complexity in exchange for high-resolution delay-Doppler estimation and more accurate interference cancellation.}

\section{Simulation Results} \label{sec:simulation_results}

\begin{table}[!t]
    \footnotesize
    \centering
    \caption{Simulation Parameters} \label{Tab:parameters}
    \begin{tabular}{c c | c c}
        \toprule
        \textbf{Parameter}                 & \textbf{Value}                 & \textbf{Parameter}             & \textbf{Value}             \\
        \midrule
        Carrier frequency $f_{\mathrm{c}}$ & \textcolor{black}{$28$~GHz}    & No. of subcarriers $N$         & $128$                      \\
        Subcarrier spacing $\Delta_f$      & \textcolor{black}{$120$~kHz}   & No. of symbols $M$             & $64$                       \\
        Symbol duration $T$                & \textcolor{black}{$8.33~\mu$s} & Noise figure $F$               & \textcolor{black}{$3$~dB}  \\
        Standard-CP duration               & \textcolor{black}{$0.59~\mu$s} & Ref. temp. $T_{\mathrm{temp}}$ & \textcolor{black}{$290$~K} \\
        Sufficient-CP duration             & \textcolor{black}{$8.33~\mu$s} & Constellation                  & 1024-QAM                   \\
        \bottomrule
    \end{tabular}
\end{table}

This section presents simulation results to validate the analytical findings from Section \ref{sec:performance_analysis} and evaluate the performance of the proposed SIC-DFT and SIC-ESPRIT algorithms developed in Section \ref{sec:SIC algorithm}. Unless otherwise noted, all simulations use the common parameter configuration summarized in Table \ref{Tab:parameters}. \textcolor{black}{For all CFAR-based methods, the detector is implemented using two-dimensional cell-averaging CFAR (CA-CFAR), with the false-alarm probability set to $P_{\mathrm{FA}}=10^{-4}$.}

    \begin{figure}[!t]
        \centering
        \includegraphics[width = 3.3 in]{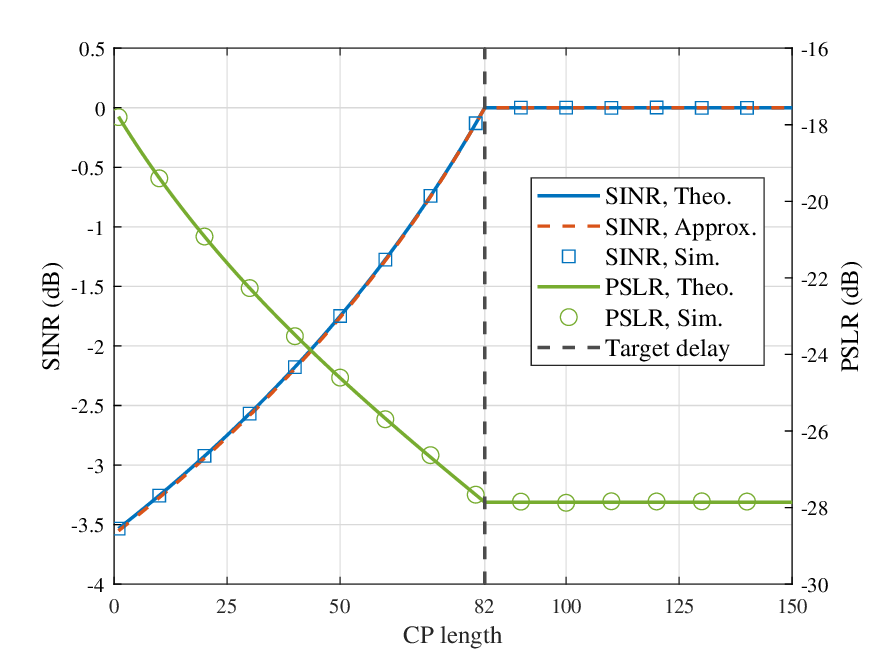}
        \vspace{-1 mm}
        \caption{SINR and PSLR versus CP length, where the target is located at $800$~m, corresponding to the $82$nd delay tap.}
        \label{fig:SINR_PSLR_CPlen}
        \vspace{-2 mm}
    \end{figure}

    \begin{figure}[!t]
        \centering
        \includegraphics[width = 3.5 in]{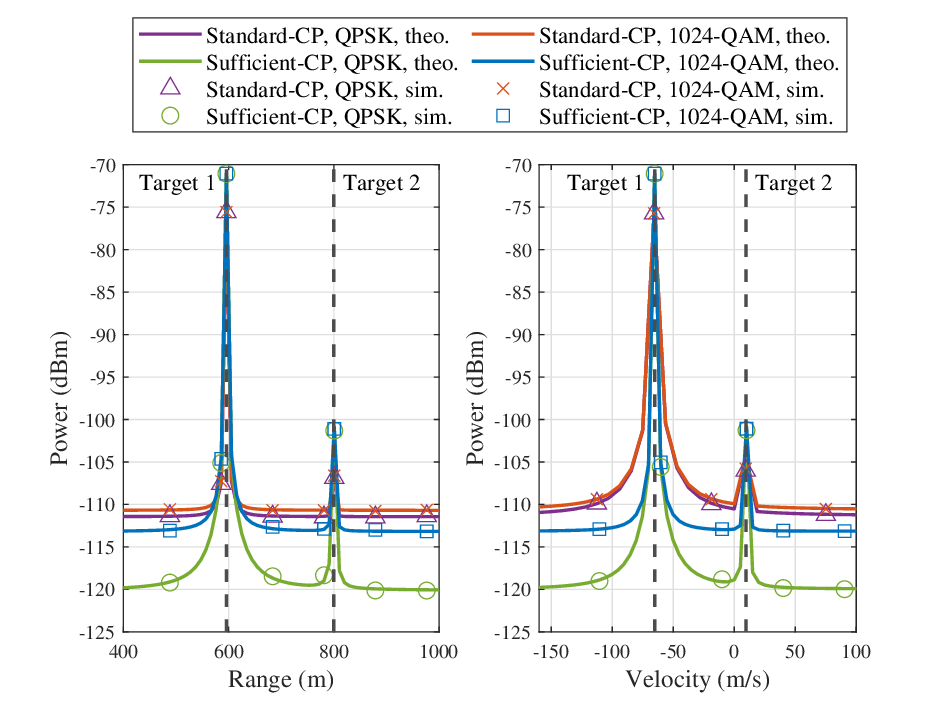}
        \vspace{-3 mm}
        \caption{Range and velocity profiles under different constellations and CP configurations.}
        \label{fig:RDM_verify}
        \vspace{-2 mm}
    \end{figure}

    \subsection{Impact of CP Length on Sensing Performance}

    We first verify the accuracy of the derived SINR/PSLR expressions and quantify the impact of the CP length on sensing performance. Fig.~\ref{fig:SINR_PSLR_CPlen} plots SINR and PSLR versus the CP length for a target at $800$~m ($82$nd delay tap) with the sensing SNR set to $0$~dB. Here, the sensing SNR of the $q$-th target is defined as $\mathrm{SNR}_q \triangleq \sigma_{\alpha,q}^{2}/\sigma^2$. The theoretical, approximate, and simulated curves closely agree, confirming the validity of our analysis. When the CP is insufficient ($N_{\mathrm{cp}}<82$), ISI/ICI arise, reducing SINR and increasing the sidelobe level; both degradations grow approximately linearly with the normalized excess delay, consistent with theory. Once $N_{\mathrm{cp}}\ge 82$, ISI/ICI vanish and performance converges to the noise-limited bound.

    Fig.~\ref{fig:RDM_verify} shows representative range/velocity profiles for different CP settings and constellations. The \emph{standard-CP} setting (i.e., normal 3GPP NR CP with $0.59~\mu$s) provides an interference-free range of $\approx 90$~m, whereas the \emph{sufficient-CP} (customized long CP with $8.33~\mu$s) baseline extends it to $\approx 1250$~m. For targets at $600$~m and $800$~m, the sufficient-CP incurs no ISI/ICI and thus represents the ideal bound. Theoretical predictions closely match the simulations, and the baseline yields lower sidelobe floors in both range and velocity. Increasing the modulation order from QPSK to 1024-QAM exacerbates sidelobe leakage, resulting in a higher sidelobe floor.

    \subsection{Performance of SIC-DFT and SIC-ESPRIT Algorithms}
\begin{figure}[!t]
        \centering
        \includegraphics[width = 3.4 in]{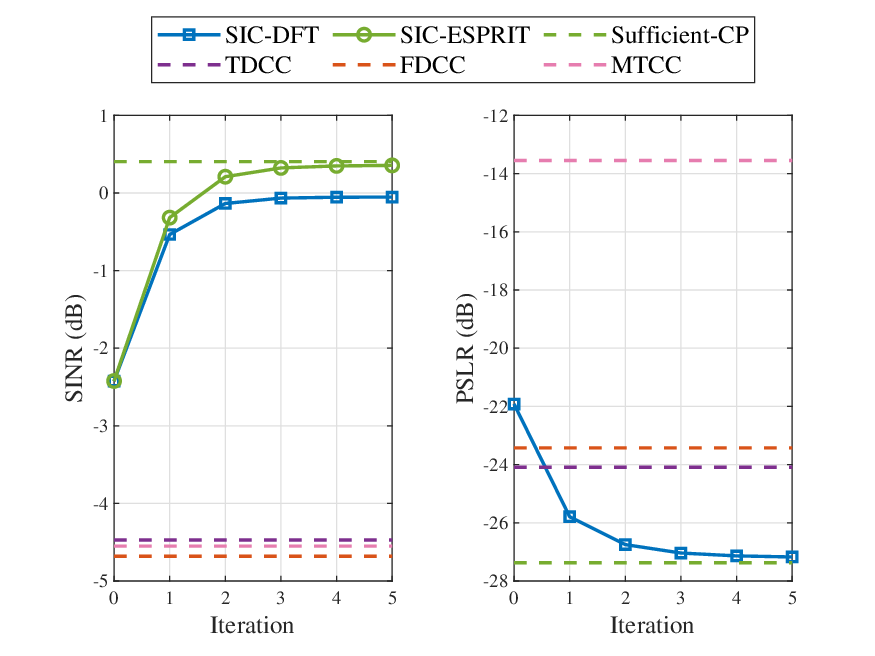}
        \vspace{-1mm}
        \caption{SINR and PSLR versus number of iterations.}
        \label{fig:SINR_PSLR_iter}
        \vspace{-2 mm}
    \end{figure}

    We next examine the proposed SIC-DFT and SIC-ESPRIT algorithms when the CP is insufficient (i.e., targets located beyond the standard-CP interference-free range). Fig.~\ref{fig:SINR_PSLR_iter} illustrates SINR and PSLR performance against the number of algorithm iterations for two representative targets located at ranges of $600$~m and $800$~m. Both targets are located beyond the standard-CP interference-free range, thus posing significant challenges due to ISI/ICI. As benchmarks, we compare against existing coherent compensation-based methods including TDCC \cite{Wang_WiOpt_2023}, \cite{Wang_TVT_2025}, FDCC \cite{Geiger_SCC_2025}, and MTCC \cite{Geiger_SCC_2025}, and include the sufficient-CP scenario as an ideal performance bound for reference. Results demonstrate that both SIC-DFT and SIC-ESPRIT rapidly converge within only a few iterations, ultimately achieving SINR gains of more than $4$~dB over the benchmark methods. SIC-ESPRIT achieves even better SINR due to its super-resolution delay-Doppler estimates, closely approaching the sufficient-CP bound. The SIC-DFT algorithm substantially reduces PSLR, also reaching the sufficient-CP bound after about~4 iterations. The SIC-ESPRIT algorithm is not included in the PSLR comparison since it does not directly generate an RDM.

    \begin{figure*}[!t]
        \centering
        \subfigbottomskip=-4pt
        \subfigcapskip=-2pt
        \subfigure[SIC-DFT with standard-CP.]{
            \includegraphics[width=0.33\linewidth]{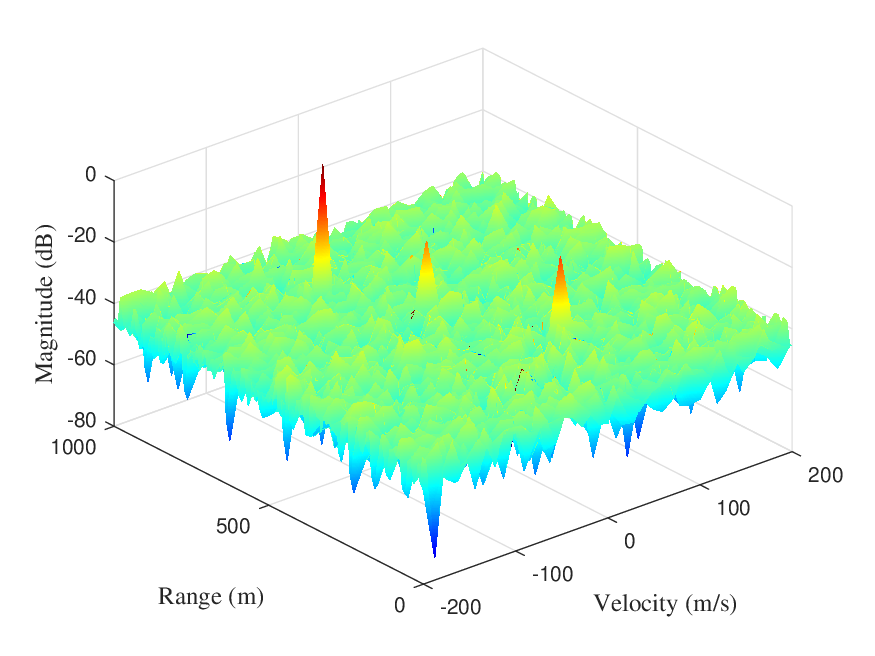} \label{fig:RDM_SICDFT}
        } \hspace{-8mm}
        \subfigure[DFT with standard-CP.]{
            \includegraphics[width=0.33\linewidth]{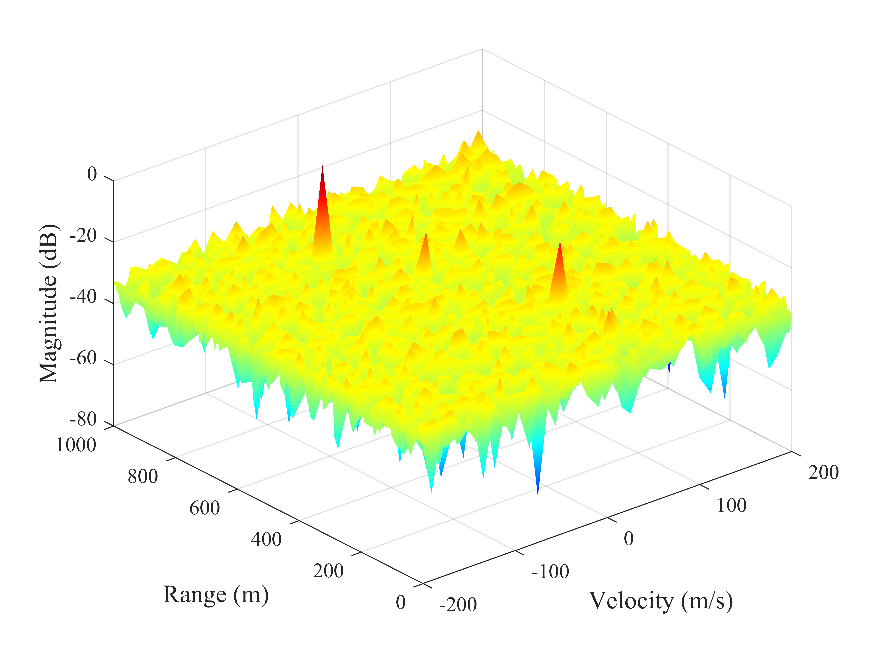} \label{fig:RDM_norCP}
        } \hspace{-8mm}
        \subfigure[DFT with sufficient-CP.]{
            \includegraphics[width=0.33\linewidth]{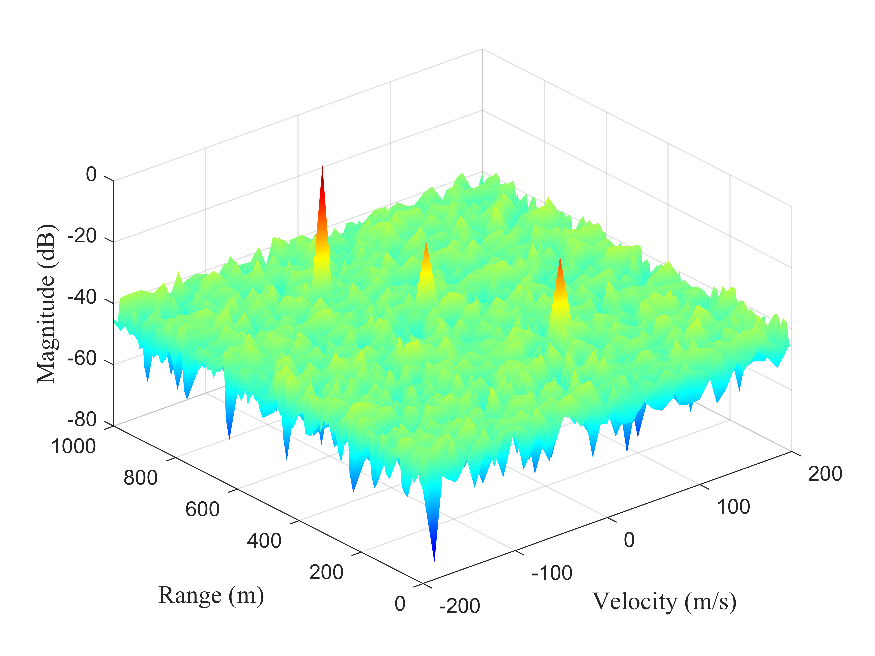} \label{fig:RDM_longCP}
        }
        \centering
        \vspace{1mm}
        \caption{RDM comparisons. }\label{fig:RDM}
        \vspace{-2mm}
    \end{figure*}

    Fig.~\ref{fig:RDM} compares representative RDMs generated by SIC-DFT under standard-CP conditions compared with conventional DFT-based processing under both standard-CP and sufficient-CP configurations. Here, three targets are present at ranges $300$~m, $600$~m, and $800$~m. Clearly, even under significant CP insufficiency, SIC-DFT dramatically reduces the sidelobe floor, nearly matching the performance achieved by the sufficient-CP baseline. This indicates that SIC-DFT can effectively reconstruct and cancel ISI/ICI, achieving high-quality target parameter estimation even with a limited CP length.

    \begin{figure}[!t]
        \centering
        \vspace{-3mm}
        \includegraphics[width = 3.4 in]{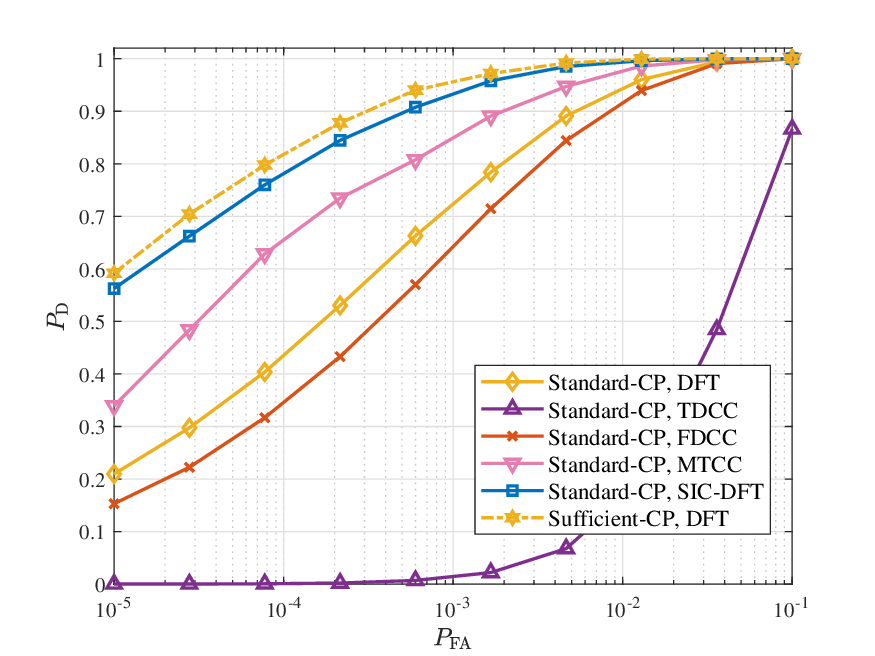}
        \vspace{-1mm}
        \caption{\textcolor{black}{Detection probability versus the prescribed CA-CFAR false-alarm probability.}}
        \label{fig:PD_PFA}
        \vspace{-4mm}
    \end{figure}

    \textcolor{black}{We also evaluate the detection performance for different prescribed CA-CFAR false-alarm probabilities. Two beyond-CP targets are located at $400$~m and $200$~m with velocities $10$~m/s and $-30$~m/s and received SNRs $15$~dB and $-10$~dB, respectively. Fig. \ref{fig:PD_PFA} plots the two-target detection probability $P_{\mathrm{D}}$ versus $P_{\mathrm{FA}}$, where a trial is counted as successful only when both targets are correctly detected within the prescribed range-Doppler tolerance. Since the prescribed $P_{\mathrm{FA}}$ associated with CA-CFAR are applied to RDMs, only RDM-based CFAR methods are included in this comparison. Increasing $P_{\mathrm{FA}}$ relaxes the CFAR threshold and improves the probability of detecting the weak target, at the cost of a higher false-alarm risk. Under the standard-CP setting, SIC-DFT consistently outperforms DFT, TDCC, FDCC, and MTCC over a broad range of $P_{\mathrm{FA}}$, and closely approaches the sufficient-CP DFT reference. This confirms that cancelling the structured CP-induced ISI/ICI improves not only the RDM sidelobe floor but also the target detection performance.}

    \begin{figure}[!t]
        \vspace{-2mm}
        \subfigcapskip = -4pt
        \centering
        \subfigure[\textcolor{black}{Range estimation RMSE.}]{
            \includegraphics[width = 3.2 in]{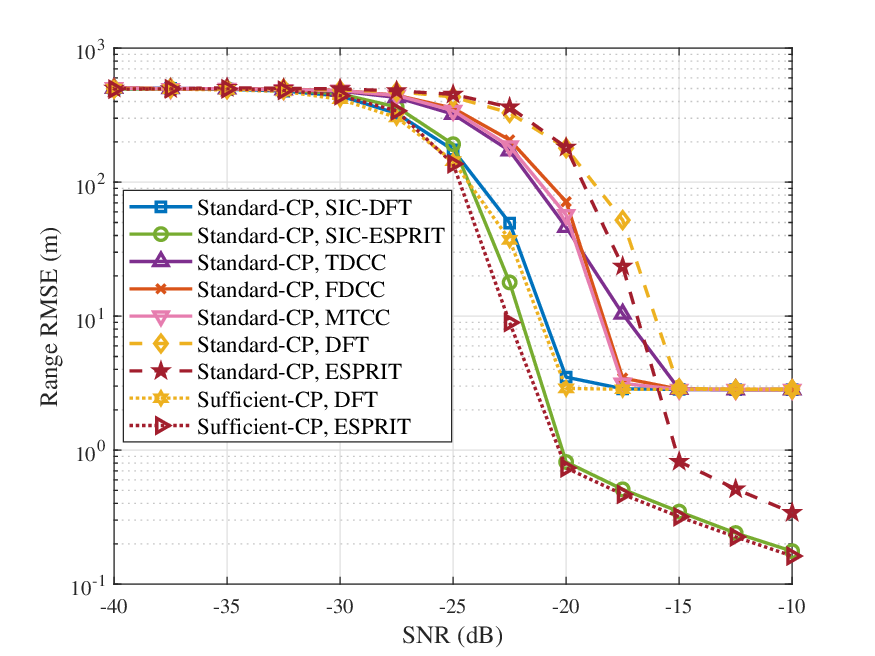}
        }\\
        \vspace{-0.2 cm}
        \subfigure[\textcolor{black}{Velocity estimation RMSE.}]{
            \includegraphics[width = 3.2 in]{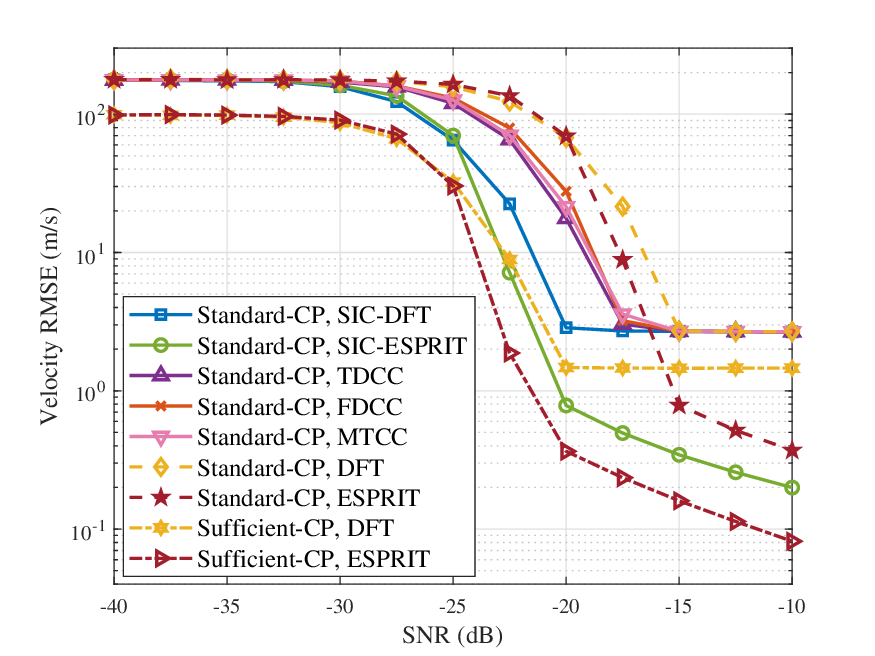}
        }
        \vspace{-0.2 cm}
        \centering
        \caption{RMSE for range and velocity estimation versus the sensing SNR.}\label{fig:RMSE_SNR}
        \vspace{-0.3 cm}
    \end{figure}

    Finally, Fig.~\ref{fig:RMSE_SNR} quantifies the range and velocity estimation accuracy by plotting RMSE against sensing SNR. \textcolor{black}{In each Monte Carlo trial, the two target ranges are uniformly drawn from the interval between $100$-$800$~m, which lies beyond the standard-CP protected range, and their velocities are uniformly chosen between $-60$ and $60$~m/s.} SIC-DFT, SIC-ESPRIT, and benchmark schemes (TDCC, FDCC, MTCC, DFT, ESPRIT) are evaluated under standard-CP conditions to assess their robustness. For comparison, we also show conventional DFT and ESPRIT results under the sufficient-CP configuration as ideal RMSE lower bounds.
    Results highlight that both SIC-DFT and SIC-ESPRIT consistently outperform all benchmarks across the entire SNR range. At moderate SNR (e.g., around $-10$~dB), SIC-ESPRIT achieves approximately one order-of-magnitude lower RMSE than the benchmarks.
    For range estimation, both SIC-DFT and SIC-ESPRIT nearly achieve the sufficient-CP ideal bound, demonstrating their ability to substantially recover lost sensing performance due to CP insufficiency. In velocity estimation, a visible gap with the bound remains since the sufficient-CP configuration benefits from a much longer observation window, enabling finer Doppler resolution. Overall, these results demonstrate that when the CP is insufficient, the proposed algorithms recover most of the sensing loss induced by ISI/ICI while remaining fully standard-compliant.

    \section{Conclusion}
    This paper has presented a unified analytical and algorithmic framework for OFDM-based ISAC systems operating beyond the CP limit. A general echo model was developed to explicitly characterize the structured ISI/ICI coupling caused by CP insufficiency. Based on this model, closed-form SINR and RDM second-order characterization, together with an approximate PSLR were derived. The analysis reveals that both SINR degradation and sidelobe elevation increase approximately linearly with the normalized excess delay beyond the CP. To mitigate these effects, two iterative interference cancellation algorithms, SIC-DFT and SIC-ESPRIT, were proposed. Simulation results validate the analytical expressions and show that both proposed algorithms can significantly improve estimation performance compared with existing benchmarks. These results offer both theoretical insight and practical techniques for achieving reliable, long-range OFDM-ISAC sensing beyond the CP limit.

    \appendices
    \section{Proof of Proposition \ref{prop:SINR}}
    \label{app:SINR_proof}
    Based on the definition of SINR in (\ref{eq:SINR_defi}), the proof proceeds by separately evaluating the expected powers of the interference-free component $\mathbb{E}\{\|\mathbf{Y}_{\mathrm{free}}\|_{F}^2\}$, the interference term $\mathbb{E}\{\| \mathbf{Y}_{\mathrm{ISI}} - \mathbf{Y}_{\mathrm{ICI}} \|_{F}^2\}$, and the noise term $\mathbb{E}\{\|\mathbf{Z}\|_{F}^2\}$.

    First, the expected power of the interference-free echo signal can be written as
    \begin{subequations}
        \begin{align}
            \mathbb{E}\{\|\mathbf{Y}_{\mathrm{free}}\|_{F}^2\}
             & = \mathbb{E}\{\mathrm{Tr}\{\mathbf{Y}_{\mathrm{free}}^H\mathbf{Y}_{\mathrm{free}}\} \}                                                                          \\
             & = \mathbb{E}\Big\{ \sum_{q=1}^Q \sum_{q'=1}^Q \alpha_q \alpha_{q'}^{\ast} \mathrm{Tr}\big\{ \mathbf{G}_q \mathbf{G}_{q'}^H \big\} \Big\} \label{eq:Yfree_deri2} \\
             & = \sum_{q=1}^Q \sigma_{\alpha,q}^{2} \mathrm{Tr}\big\{\mathbb{E}\{ \mathbf{G}_q \mathbf{G}_q^H \} \big\},
        \end{align}
    \end{subequations}
    where we define $\mathbf{G}_q \triangleq \mathbf{b}(\tau_q)\mathbf{c}^H(f_{\mathrm{d},q}) \odot \mathbf{S}$ for brevity. \textcolor{black}{The cross terms in (\ref{eq:Yfree_deri2}) vanish because the target reflection coefficients are independent and zero-mean, i.e., $\mathbb{E}\{\alpha_q\alpha_{q'}^{\ast}\}=0$ for $q\neq q'$. The diagonal terms are given by  $\mathbb{E}\{|\alpha_q|^2\}=\sigma_{\alpha,q}^{2}$.} Since the transmitted data symbols $s_{n,m}$ are i.i.d. with unit power, $\mathbb{E}\{\mathbf{G}_q \mathbf{G}_q^H\}$ can be derived as
    \begin{subequations}
        \begin{align}
              & \sum_{m=0}^{M-1} \big|[\mathbf{c}(f_{\mathrm{d},q})]_m\big|^2 \mathbb{E}\big\{ (\mathbf{b}(\tau_q)\odot \mathbf{s}_m)  (\mathbf{b}(\tau_q)\odot \mathbf{s}_m)^H \big\} \\
            = & \sum_{m=0}^{M-1} \mathrm{diag}\{\mathbf{b}(\tau_q)\} \mathbb{E}\{\mathbf{s}_m \mathbf{s}_m^H\} \mathrm{diag}\{\mathbf{b}^H(\tau_q)\}                                   \\
            = & M \mathbf{I}_N. \label{eq:expe_GG3}
        \end{align}
    \end{subequations}
    Thus, the expected power of the interference-free component is given by
    \begin{equation}
        \mathbb{E}\{\|\mathbf{Y}_{\mathrm{free}}\|_{F}^2\} = MN \sum_{q=1}^Q \sigma_{\alpha,q}^{2}.
    \end{equation}

    Next, we evaluate the interference power, which consists of the ISI, ICI, and their cross term:
    \begin{subequations}
        \begin{align}
              & \mathbb{E}\{\| \mathbf{Y}_{\mathrm{ISI}} - \mathbf{Y}_{\mathrm{ICI}} \|_{F}^2\} \notag                                                                                                                                 \\
            = & \mathbb{E}\big\{ \mathrm{Tr}\{ (\mathbf{Y}_{\mathrm{ISI}} - \mathbf{Y}_{\mathrm{ICI}})(\mathbf{Y}_{\mathrm{ISI}} - \mathbf{Y}_{\mathrm{ICI}})^H \} \big\}                                                              \\
            = & \mathbb{E}\{\|\mathbf{Y}_{\mathrm{ISI}}\|_{F}^2\} + \mathbb{E}\{\|\mathbf{Y}_{\mathrm{ICI}}\|_{F}^2\} - 2\mathfrak{R}\big\{\mathbb{E}\{ \mathrm{Tr}\{\mathbf{Y}^H_{\mathrm{ISI}}\mathbf{Y}_{\mathrm{ICI}} \} \}\big\}.
        \end{align}
    \end{subequations}
    For the ICI component, the expected power is expressed as
    \begin{subequations}
        \begin{align}
            \mathbb{E}\{\|\mathbf{Y}_{\mathrm{ICI}}\|_{F}^2\}
             & = \sum_{q=\tilde{Q}+1}^Q \sigma_{\alpha,q}^{2} \mathbb{E}\big\{ \mathrm{Tr}\{ \mathbf{G}_q^H \mathbf{\Phi}_q^H \mathbf{\Phi}_q \mathbf{G}_q \} \big\}                    \\
             & = \sum_{q=\tilde{Q}+1}^Q \sigma_{\alpha,q}^{2} \mathrm{Tr}\big\{\mathbf{\Phi}_q^H \mathbf{\Phi}_q \mathbb{E}\{ \mathbf{G}_q \mathbf{G}_q^H\} \big\}. \label{eq:ICI_pow2}
        \end{align}
    \end{subequations}
    Using the derivations in (\ref{eq:expe_GG3}), the key step reduces to evaluating $\mathrm{Tr}\big\{\mathbf{\Phi}_q \mathbf{\Phi}_q^H\big\}$, which can be expanded as
    \begin{subequations} \label{eq:phi_sum}
        \begin{align}
            \mathrm{Tr}\big\{\mathbf{\Phi}_q \mathbf{\Phi}_q^H\big\}
             & = \sum_{n, n'=0}^{N-1} \Big|\frac{1}{N} \sum_{i=0}^{l_q-N_{\mathrm{cp}}-1} e^{\jmath \frac{2\pi}{N}(n'-n)i} \Big|^2 \\
             & = \frac{1}{N^2} \sum_{n, n'=0}^{N-1} \sum_{i,i'=0}^{l_q-N_{\mathrm{cp}}-1} e^{\jmath \frac{2\pi}{N}(n'-n)(i-i')}    \\
             & = \frac{1}{N} \sum_{i,i'=0}^{l_q-N_{\mathrm{cp}}-1} \sum_{\Delta=0}^{N-1} e^{\jmath \frac{2\pi}{N}\Delta(i-i')}     \\
             & = l_q - N_{\mathrm{cp}}, \label{eq:Trace_phi}
        \end{align}
    \end{subequations}
    where $\delta_{n,n'}$ denotes the Kronecker delta function. Substituting (\ref{eq:expe_GG3}) and (\ref{eq:Trace_phi}) into (\ref{eq:ICI_pow2}) yields
    \begin{equation}
        \mathbb{E}\{\|\mathbf{Y}_{\mathrm{ICI}}\|_{F}^2\} = M \sum_{q=\tilde{Q}+1}^Q (l_q - N_{\mathrm{cp}}) \sigma_{\alpha,q}^{2}.
    \end{equation}

    Similarly, the expectation of the ISI power is written as
    \begin{subequations}
        \begin{align}
            \!\!\!\!\!\mathbb{E}\{\|\mathbf{Y}_{\mathrm{ISI}}\|_{F}^2\}
             & \!= \!\!\!\sum_{q=\tilde{Q}+1}^Q\!\! \sigma_{\alpha,q}^{2} \mathbb{E}\big\{ \mathrm{Tr}\{\mathbf{J}_1^H \tilde{\mathbf{G}}_q^H \mathbf{\Phi}_q^H \mathbf{\Phi}_q \tilde{\mathbf{G}}_q \mathbf{J}_1 \}\! \big\}  \!\!\!\! \\
             & \!= \!\!\!\sum_{q=\tilde{Q}+1}^Q\!\! \sigma_{\alpha,q}^{2} \mathrm{Tr}\big\{\mathbf{\Phi}_q^H \mathbf{\Phi}_q \mathbb{E}\{ \tilde{\mathbf{G}}_q \mathbf{J}_1 \mathbf{J}_1^H \tilde{\mathbf{G}}_q^H\}\! \big\}, \!\!\!
        \end{align}
    \end{subequations}
    where $\tilde{\mathbf{G}}_q \triangleq \mathbf{b}(\tau_q-T_{\mathrm{cp}})\mathbf{c}^H(f_{\mathrm{d},q}) \odot \mathbf{S}$. Since $\mathbf{J}_1 \mathbf{J}_1^H = \mathrm{diag}\{1, \dots, 1, 0\} \in\mathbb{C}^{M \times M}$, the expectation of $ \tilde{\mathbf{G}}_q \mathbf{J}_1 \mathbf{J}_1^H \tilde{\mathbf{G}}_q^H$ can be expanded as
    \begin{equation}
        \begin{aligned}
              & \!\sum_{m=0}^{M-2} \! \big|[\mathbf{c}(f_{\mathrm{d},q})]_m\big|^2 \mathbb{E}\big\{ (\mathbf{b}(\tau_q \!-\! T_{\mathrm{cp}})\!\odot\! \mathbf{s}_m)(\mathbf{b}(\tau_q \!-\! T_{\mathrm{cp}}) \!\odot\! \mathbf{s}_m)^H \!\big\} \!\! \\
            = & (M-1) \mathbf{I}_N,
        \end{aligned}
    \end{equation}
    which implies
    \begin{equation}
        \mathbb{E}\{\|\mathbf{Y}_{\mathrm{ISI}}\|_{F}^2\} = (M-1) \sum_{q=\tilde{Q}+1}^Q (l_q - N_{\mathrm{cp}}) \sigma_{\alpha,q}^{2}.
    \end{equation}

    Then, we derive the cross-correlation term between the ISI and ICI components:
    \begin{equation}
        \!\!\!\!\! \mathbb{E}\{ \mathrm{Tr}\{\mathbf{Y}^H_{\mathrm{ISI}}  \mathbf{Y}_{\mathrm{ICI}} \} \} \! = \!\!\! \sum_{q=\tilde{Q}+1}^Q \!\!\!\! \sigma_{\alpha,q}^{2} \mathrm{Tr}\big\{\! \mathbf{\Phi}_q^H \! \mathbf{\Phi}_{q} \mathbb{E}\{\mathbf{G}_q\mathbf{J}_1^H \! \tilde{\mathbf{G}}_q^H \} \!\big\}. \!\!\!\!
    \end{equation}
    Note that $\mathbb{E}\{\mathbf{G}_q\mathbf{J}_1^H \tilde{\mathbf{G}}_q^H\}$ can be derived as
    \begin{subequations}
        \begin{align}
              & \sum_{m,m'} \mathbb{E}\{(\mathbf{b}(\tau_q) \odot \mathbf{s}_m)(\mathbf{b}(\tau_q-T_{\mathrm{cp}}) \odot \mathbf{s}_{m'})^H\}          \notag                 \\
              & ~~~~~ \times e^{\jmath 2 \pi (m-m') f_{\mathrm{d},q} T_{\mathrm{s}}} [\mathbf{J}_1^H]_{m,m'}                                                                  \\
            = & \sum_{m,m'} \delta_{m,m'} \mathrm{diag}\{\mathbf{b}(T_{\mathrm{cp}})\} e^{\jmath 2 \pi (m-m') f_{\mathrm{d},q} T_{\mathrm{s}}} [\mathbf{J}_1^H]_{m,m'} \!\!\! \\
            = & \sum_{m=0}^{M-1} [\mathbf{J}_1^H]_{m,m} \mathrm{diag}\{\mathbf{b}(T_{\mathrm{cp}})\}.
        \end{align}
    \end{subequations}
    Since all diagonal entries of $\mathbf{J}_1^H$ are zero, we obtain
    \begin{equation} \label{eq:cross_ISI_ICI}
        \mathbb{E}\{ \mathrm{Tr}\{\mathbf{Y}^H_{\mathrm{ISI}}\mathbf{Y}_{\mathrm{ICI}} \} \} = 0,
    \end{equation}
    which indicates that the ISI and ICI components are statistically uncorrelated, allowing their powers to be added directly to form the total interference power.

    In addition, the noise power is given by
    \begin{equation}
        \mathbb{E}\{\|\mathbf{Z}\|_F^2\} = MN \sigma^2.
    \end{equation}
    Combining the above results, the SINR of the echo signal can be written as
    \begin{equation}
        \textcolor{black}{\mathrm{SINR} =\frac{MN\sum_{q=1}^Q \sigma_{\alpha,q}^{2}}{(2M-1)N \sum_{q=\tilde{Q}+1}^Q \rho_q \sigma_{\alpha,q}^{2} + MN\sigma^2}.}
    \end{equation}

    \section{Proof of Proposition \ref{prop:RDM}}
    \label{app:RDM_proof}
    Since the sensing observations consist of the interference-free components, i.e., the CP-induced ISI/ICI components and AWGN, the RDM in \eqref{eq:RDM} can be decomposed as
    \begin{equation}
        \boldsymbol{\chi} = \boldsymbol{\chi}_{\mathrm{free}} + \boldsymbol{\chi}_{\mathrm{ISI}} - \boldsymbol{\chi}_{\mathrm{ICI}} + \boldsymbol{\chi}_{\mathrm{awgn}},
    \end{equation}
    where the four terms are given by
    \begin{subequations}
        \begin{align}
            \chi_{\mathrm{free}} (l, \nu) & = \sum_{q=1}^Q \frac{\alpha_q}{\sqrt{MN}} \sum_{m,n} |s_{n,m}|^2 e^{\jmath \theta_{n,m,q}},                                                    \\
            \chi_{\mathrm{ISI}} (l, \nu)  & = \!\! \sum_{q=\tilde{Q}+1}^Q \! \frac{\alpha_q}{\sqrt{MN}} \!\! \sum_{m,n,n'} \! s_{n',m-1} s_{n,m}^{\ast} \psi^{\mathrm{ISI}}_{n,n',m}, \!\! \\
            \chi_{\mathrm{ICI}} (l, \nu)  & = \!\! \sum_{q=\tilde{Q}+1}^Q \!\! \frac{\alpha_q}{\sqrt{MN}} \sum_{m,n,n'} s_{n',m} s_{n,m}^{\ast} \psi^{\mathrm{ICI}}_{n,n',m},              \\
            \chi_{\mathrm{awgn}}(l,\nu)   & = \frac{1}{\sqrt{MN}} \sum_{m,n} z_{n,m} s_{n,m}^{\ast} e^{\jmath \frac{2\pi}{N} nl} e^{-\jmath \frac{2\pi}{M} m \nu},                         \\
            \theta_{n,m,q}                & = \frac{2\pi}{N} n(l-\tilde{l}_q) + \frac{2\pi}{M} m (\tilde{\nu}_q - \nu),                                                                    \\
            \psi^{\mathrm{ISI}}_{n,n',m}  & = \phi^q_{n,n'} e^{\jmath \frac{2\pi}{N} (nl - n'(\tilde{l}_q-N_{\mathrm{cp}}) )} e^{\jmath \frac{2\pi}{M}  ((m-1)\tilde{\nu}_q - m\nu) },     \\
            \psi^{\mathrm{ICI}}_{n,n',m}  & = \phi^q_{n,n'} e^{\jmath \frac{2\pi}{N} (nl - n'\tilde{l}_q)} e^{\jmath \frac{2\pi}{M} m (\tilde{\nu}_q - \nu) }.
        \end{align}
    \end{subequations}
    Thus, the second-order moment of the RDM is written as
    \begin{equation} \label{eq:RDM_expect}
        \begin{aligned}
             & \mathbb{E}\big\{ |\chi(l,\nu)|^2 \big\} = \mathbb{E}\big\{|\chi_{\mathrm{free}}(l,\nu)|^2\big\} + \mathbb{E}\big\{|\chi_{\mathrm{ISI}}(l,\nu)|^2\big\}                                                  \\
             & + \! \mathbb{E}\big\{|\chi_{\mathrm{ICI}}(l,\nu)|^2\!\big\}  \!+\! \mathbb{E}\big\{|\chi_{\mathrm{awgn}}(l,\nu)|^2\!\big\} \!+\! 2 \mathfrak{R} \Big\{ \! \mathbb{E}\big\{ \!\chi_{\mathrm{sig}}(l,\nu) \\
             & (\chi_{\mathrm{awgn}}(l,\nu))^{\ast} \big\}  + \mathbb{E}\big\{ \chi_{\mathrm{free}}(l,\nu) (\chi_{\mathrm{ISI}}(l,\nu))^{\ast} \big\} - \mathbb{E}\big\{                                               \\
             & \chi_{\mathrm{free}}(l,\nu) (\chi_{\mathrm{ICI}}(l,\nu))^{\ast} \big\} - \mathbb{E}\big\{ \chi_{\mathrm{ISI}}(l,\nu) (\chi_{\mathrm{ICI}}(l,\nu))^{\ast} \big\} \Big\},
        \end{aligned}
    \end{equation}
    where $\boldsymbol{\chi}_{\mathrm{sig}} = \boldsymbol{\chi}_{\mathrm{free}} + \boldsymbol{\chi}_{\mathrm{ISI}} - \boldsymbol{\chi}_{\mathrm{ICI}}$, and the last term collects the cross-correlations between the interference-free signal, ISI, ICI, and noise components.

    In the subsequent derivations, we will repeatedly encounter fourth-order moments of the modulated symbols. \textcolor{black}{Let $s_i,s_j,s_k,s_l$ denote four data symbols indexed over the time-frequency grid. Under the i.i.d. normalized-symbol assumptions $\mathbb{E}\{s_{n,m}\} = \mathbb{E}\{s_{n,m}^2 \} = 0$, $\mathbb{E}\{|s_{n,m}|^2\} = 1$, and $\mu_4=\mathbb{E}\{|s_{n,m}|^4\}$, the mixed fourth-order moment is nonzero only for the two conjugate pairings $(i=j,k=l)$ and  $(i=l,j=k)$. When all four indices coincide, the fourth-order moment equals $\mu_4$, which requires a correction to avoid double counting. Therefore,}
    \begin{equation} \label{eq:expectation_symbol}
        \textcolor{black}{\mathbb{E}\{s_i s_j^{\ast} s_k s_l^{\ast}\} = \delta_{ij}\delta_{kl}+\delta_{il}\delta_{jk}+(\mu_4-2)\delta_{ij}\delta_{jk}\delta_{kl}.}
    \end{equation}
    We now analyze each term in (\ref{eq:RDM_expect}).
    First, the second-order moment of the interference-free component is
    \begin{subequations}
        \begin{align}
              & \mathbb{E}\big\{ |\chi_{\mathrm{free}}(l,\nu)|^2 \big\} \notag                                                                                                            \\
            = & \sum_{q = 1}^Q \frac{\sigma_{\alpha,q}^{2}}{MN} \sum_{m,n} \sum_{m',n'} \mathbb{E}\{|s_{n,m}|^2 |s_{n',m'}|^2\} e^{\jmath (\theta_{n,m,q} - \theta_{n',m',q})}            \\
            = & \sum_{q = 1}^Q \frac{\sigma_{\alpha,q}^{2}}{MN} \!\!\!\! \sum_{\substack{m,n                                                                                              \\(m',n') \neq (m,n)}} \!\!\!\! e^{\jmath (\theta_{n,m,q}-\theta_{n',m',q})} +  \mu_4 \sum_{q = 1}^Q \sigma_{\alpha,q}^{2} \!\! \\
            = & \sum_{q = 1}^Q \frac{\sigma_{\alpha,q}^{2}}{MN} \Big( \Big|\sum_{m,n} e^{\jmath \theta_{n,m,q}}\Big|^2 - MN\Big) + \mu_4 \sum_{q = 1}^Q \sigma_{\alpha,q}^{2}             \\
            = & \sum_{q = 1}^Q \frac{\sigma_{\alpha,q}^{2}}{MN} |D_N(l \!-\! \tilde{l}_q)|^2 |D_M(\nu \!-\! \tilde{\nu}_q)|^2 \!+\! (\mu_4 \!-\! 1) \sum_{q = 1}^Q \sigma_{\alpha,q}^{2}.
        \end{align}
    \end{subequations}

    Next, the second-order moment of the ISI contribution can be expressed as
    \begin{subequations}
        \begin{align}
            \mathbb{E}\big\{ |\chi_{\mathrm{ISI}}(l,\nu)|^2 \big\}
             & = \!\sum_{q=\tilde{Q}+1}^Q\!\! \frac{\sigma_{\alpha,q}^{2}}{MN} \!\sum_{m = 1}^{M-1}\! \sum_{n=0}^{N-1}\! \sum_{n'=0}^{N-1}\! |\psi^{\mathrm{ISI}}_{n,n',m}|^2       \\
             & = \!\sum_{q=\tilde{Q}+1}^Q \!\!\! \frac{\sigma_{\alpha,q}^{2}(M-1)}{MN} \!\sum_{n=0}^{N-1} \!\sum_{n'=0}^{N-1} \! | \phi_{n,n'}^{q} |^2   \label{eq:RDM_ISI_c}  \!\! \\
             & \textcolor{black}{\approx} \sum_{q=\tilde{Q}+1}^Q \rho_q \sigma_{\alpha,q}^{2},
        \end{align}
    \end{subequations}
    where (\ref{eq:RDM_ISI_c}) follows from (\ref{eq:phi_sum}). \textcolor{black}{For finite-frame processing, the out-of-frame symbol is set to zero, i.e., $s_{n,-1} = 0$. Hence, the ISI summation is strictly over $m=1,\ldots,M-1$. This finite-frame boundary introduces an $\mathcal{O}(1/M)$ correction. For compactness, we neglect this single-symbol boundary correction in the subsequent second-order RDM analysis.} Similarly, the second-order moment of the ICI contribution is given by
    \begin{subequations}
        \begin{align}
              & \mathbb{E}\big\{ |\chi_{\mathrm{ICI}}(l,\nu)|^2 \big\}   \notag                                                                                                                                                         \\
            = & \sum_{q=\tilde{Q}+1}^Q \frac{\sigma_{\alpha,q}^{2}}{MN} \sum_{m_1,n_1} \sum_{(m_2,n_2) \neq (m_1,n_1)} \rho_q^2 e^{\jmath(\theta_{n_1,m_1,q} - \theta_{n_2,m_2,q})} \notag                                              \\
              & + \!\!\sum_{q=\tilde{Q}+1}^Q \!\! \frac{\sigma_{\alpha,q}^{2}}{MN} \!\Big( \!\!\! \sum_{m,n, n' \neq n} \!\!\! |\psi^{\mathrm{ICI}}_{n,n',m}|^2 \!+\! \mu_4 \sum_{m,n} |\rho_q e^{\jmath \theta_{n,m,q}} |^2 \Big) \!\! \\
            = & \sum_{q=\tilde{Q}+1}^Q \!\! \frac{\rho_q^2 \sigma_{\alpha,q}^{2}}{MN} \big| \sum_{m,n} e^{\jmath \theta_{n,m,q}} \big|^2 \!+\! (\rho_q \!+\! (\mu_4 \!-\! 2) \rho_q^2) \sigma_{\alpha,q}^{2} \!\!\!                     \\
            = & \sum_{q=\tilde{Q}+1}^Q \frac{\rho_q^2 \sigma_{\alpha,q}^{2}}{MN} |D_N(l - \tilde{l}_q)|^2 |D_M(\nu - \tilde{\nu}_q)|^2 \notag                                                                                           \\
              & + \sum_{q=\tilde{Q}+1}^Q \big((\mu_4 - 1) \rho_q^2 + \rho_q(1-\rho_q) \big) \sigma_{\alpha,q}^{2}.
        \end{align}
    \end{subequations}
    Finally, the noise contribution is \vspace{-1mm}
    \begin{subequations}
        \begin{align}
            \mathbb{E}\big\{ |\chi_{\mathrm{awgn}}(l,\nu)|^2 \big\}
            = & \frac{\sigma^2}{MN} \sum_{m=0}^{M-1} \sum_{n=0}^{N-1} \mathbb{E}\{|s_{n,m}|^2\} \\
            = & \sigma^2 .
        \end{align}
    \end{subequations}

    Having derived the auto-correlation terms, we proceed to examine the cross-correlation terms between different components in (\ref{eq:RDM_expect}).
    Specifically, the cross-correlation between $\chi_{\mathrm{sig}}(l,\nu)$ and $\chi_{\mathrm{awgn}}(l,\nu)$ can be expressed as \vspace{-1mm}
    \begin{subequations}
        \begin{align}
             & \mathbb{E}\big\{ \chi_{\mathrm{sig}}(l,\nu) (\chi_{\mathrm{awgn}}(l,\nu))^{\ast} \big\} = \frac{1}{MN} \mathbb{E}_{\alpha, s}\Big\{ \chi_{\mathrm{sig}}(l,\nu)\notag \\
             & \hspace{1.5cm} \times \sum_{m,n} \mathbb{E}_{z}\{z_{n,m}^{\ast}\} s_{n,m} e^{-\jmath \frac{2\pi}{N}nl} e^{\jmath \frac{2\pi}{M}m\nu} \Big\}                          \\
             & \hspace{4.0cm} =  ~ 0.
        \end{align}
    \end{subequations}
    The cross-correlation between the interference-free and ISI components is \vspace{-1mm}
    \begin{subequations}
        \begin{align}
             & \mathbb{E}\big\{ \chi_{\mathrm{free}}(l,\nu) (\chi_{\mathrm{ISI}}(l,\nu))^{\ast} \big\} = \! \sum_{q=\tilde{Q}+1}^Q \!\!\! \frac{\sigma_{\alpha,q}^{2}}{MN} \!\!\! \sum_{\substack{m_1,n_1 \\m_2,n_2,n'_2}} \!\!\! \mathbb{E}\big\{ |s_{n_1,m_1}|^2  \notag \\
             & \hspace{1.5cm} \times s_{n'_2, m_2-1}^{\ast} s_{n_2, m_2} \big\} e^{\jmath \theta_{n_1,m_1,q}} (\psi^{\mathrm{ISI}}_{n_2,n'_2,m_2})^{\ast}                                                 \\
             & \hspace{3.8cm} = 0.
        \end{align}
    \end{subequations}
    Similarly, the cross-correlation between $\chi_{\mathrm{free}}(l,\nu)$ and $\chi_{\mathrm{ICI}}(l,\nu)$ is \vspace{-1mm}
    \begin{subequations}
        \begin{align}
              & \mathbb{E}\big\{ \chi_{\mathrm{free}}(l,\nu) (\chi_{\mathrm{ICI}}(l,\nu))^{\ast} \big\} \!= \!\!\!\sum_{q=\tilde{Q}+1}^Q \!\!\! \frac{\sigma_{\alpha,q}^{2}}{MN} \!\!\sum_{m_1,n_1}\! \sum_{(m_2,n_2) \neq (m_1,n_1)}    \notag \\
              & \hspace{1.5 cm} \rho_q e^{\jmath(\theta_{n_1,m_1,q} - \theta_{n_2,m_2,q})}  + \sum_{m,n} \mu_4 \rho_q \sigma_{\alpha,q}^{2}                                                                                                     \\
            = & \!\! \sum_{q=\tilde{Q}+1}^Q \!\!\!\! \frac{\rho_q \sigma_{\alpha,q}^{2}}{MN} |D_N(l \!-\! \tilde{l}_q)|^2 |D_M(\nu \!-\! \tilde{\nu}_q)|^2 \!+\! (\mu_4 -1) \rho_q \sigma_{\alpha,q}^{2}.
        \end{align}
    \end{subequations}
    Finally, the cross-correlation between $\chi_{\mathrm{ISI}}(l,\nu)$ and $\chi_{\mathrm{ICI}}(l,\nu)$ can be written as \vspace{-1mm}
    \begin{subequations}
        \begin{align}
             & \hspace{-0.3cm} \mathbb{E}\big\{ \chi_{\mathrm{ISI}}(l,\nu) (\chi_{\mathrm{ICI}}(l,\nu))^{\ast} \big\} = \sum_{q=\tilde{Q}+1}^Q  \frac{\sigma_{\alpha,q}^{2}}{MN}  \sum_{\substack{m_1,n_1,n'_1 \\m_2,n_2,n'_2}}  \notag \\
             & \mathbb{E}\{ s_{n'_1,m_1-1} s_{n_1,m_1}^{\ast} s_{n'_2,m_2}^{\ast}  s_{n_2, m_2} \} \psi^{\mathrm{ISI}}_{n_1, n'_1, m_1} \psi^{\mathrm{ICI}}_{n_2, n'_2, m_2}                                   \\
             & \hspace{3.5 cm} = 0.
        \end{align}
    \end{subequations}
    Based on the above derivations, the second-order moment of the RDM is given by
    \begin{equation}
        \mathbb{E}\big\{ |\chi(l,\nu)|^2 \big\} = \sum_{q = 1}^Q \frac{\tilde{\sigma}_{\alpha,q}^{2}}{MN} |D_N(l \!-\! \tilde{l}_q)|^2 |D_M(\nu \!-\! \tilde{\nu}_q)|^2 + \sigma_{\mathrm{SL}}^2.
    \end{equation}

    \section{Proof of Proposition \ref{prop:cov_matrix}}
    \label{app:covMatr_proof}
    To derive the covariance matrix of $\hat{\mathbf{h}}$, we first decompose the $i$-th entry of $\hat{\mathbf{h}}$ as
    \begin{equation}
        \hat{h}_i = \hat{h}_i^{\mathrm{free}} + \hat{h}_i^{\mathrm{ISI}} - \hat{h}_i^{\mathrm{ICI}} + \hat{h}_i^{\mathrm{awgn}},
    \end{equation}
    where $\hat{h}_i^{\mathrm{free}} = y_{n,m}^{\mathrm{free}} s_{n,m}^{\ast}$, $\hat{h}_i^{\mathrm{ISI}} = y_{n,m}^{\mathrm{ISI}} s_{n,m}^{\ast}$, $\hat{h}_i^{\mathrm{ICI}} = y_{n,m}^{\mathrm{ICI}} s_{n,m}^{\ast}$, $\hat{h}_i^{\mathrm{awgn}} = z_{n,m} s_{n,m}^{\ast}$, and $i = n + m N$. Accordingly, the $(i,j)$-th entry of the covariance matrix $\mathbf{R}$ can be written as
    \begin{equation} \label{eq:cov_entry}
        \begin{aligned}
             & [\mathbf{R}]_{i,j} = \mathbb{E}\{ \hat{h}_{i}^{\mathrm{free}} (\hat{h}_{j}^{\mathrm{free}})^{\ast} \} \!+\! \mathbb{E}\{ \hat{h}_{i}^{\mathrm{ISI}} (\hat{h}_{j}^{\mathrm{ISI}})^{\ast} \} \!+\! \mathbb{E}\{ \hat{h}_{i}^{\mathrm{ICI}} (\hat{h}_{j}^{\mathrm{ICI}})^{\ast} \}                      \\
             & ~ + \! \mathbb{E}\{ \hat{h}_{i}^{\mathrm{awgn}} (\hat{h}_{j}^{\mathrm{awgn}})^{\ast} \!\} \!+\! 2\mathfrak{R}\big\{ \mathbb{E}\{ \hat{h}_{i}^{\mathrm{free}} (\hat{h}_{j}^{\mathrm{ISI}})^{\ast} \!\} \!-\! \mathbb{E}\{ \hat{h}_{i}^{\mathrm{free}} (\hat{h}_{j}^{\mathrm{ICI}})^{\ast} \!\} \!\!\! \\
             & ~  - \mathbb{E}\{ \hat{h}_{i}^{\mathrm{ISI}} (\hat{h}_{j}^{\mathrm{ICI}})^{\ast} \} + \mathbb{E}\{ (\hat{h}_{i}^{\mathrm{free}} + \hat{h}_{i}^{\mathrm{ISI}} - \hat{h}_{i}^{\mathrm{ICI}}) (\hat{h}_{j}^{\mathrm{awgn}})^{\ast} \}  \big\}.
        \end{aligned}
    \end{equation}

    We now evaluate each term in (\ref{eq:cov_entry}). Based on the signal model in (\ref{eq:sig_freq}), the covariance of the interference-free component $\hat{h}_{i}^{\mathrm{free}}$ is
    \begin{subequations}
        \begin{align}
            \mathbb{E}\{ \hat{h}_{i}^{\mathrm{free}} (\hat{h}_{j}^{\mathrm{free}})^{\ast} \}
             & = \sum_{q=1}^Q \sigma_{\alpha,q}^{2} \mathbb{E}\{ |s_{n,m}|^2 |s_{n',m'}|^2 \!\} [\mathbf{a}_q]_{i} [\mathbf{a}_q^{\ast}]_{j} \\
             & = \begin{cases}
                     \mu_4 \sum_{q=1}^Q \sigma_{\alpha,q}^{2}, ~                                        & i = j;    \\
                     \sum_{q=1}^Q \sigma_{\alpha,q}^{2} [\mathbf{a}_q]_{i} [\mathbf{a}_q^{\ast}]_{j}, ~ & i \neq j.
                 \end{cases}
        \end{align}
    \end{subequations}
    The covariance of the ISI component $\hat{h}_{i}^{\mathrm{ISI}}$ can be obtained by
    \begin{subequations}
        \begin{align}
                    & \mathbb{E}\{ \hat{h}_{i}^{\mathrm{ISI}} (\hat{h}_{j}^{\mathrm{ISI}})^{\ast} \} \notag                                                                         \\
            =       & \sum_{q=\tilde{Q}+1}^Q \sigma_{\alpha,q}^{2} \sum_{n_1, n_2} \mathbb{E}\{ s_{n_1,m-1} s_{n,m}^{\ast} s_{n_2, m'-1}^{\ast} s_{n',m'} \} \phi_{n,n_1}^q  \notag \\
                    & \times (\phi_{n',n_2}^q)^{\ast} e^{\jmath 2 \pi (n_1 -n_2)\Delta_f (T_{\mathrm{cp}}-\tau_q)} e^{\jmath 2 \pi (m-m') f_{\mathrm{d},q} T_{\mathrm{s}}}          \\
            \approx & \begin{cases}
                          \sum_{q=\tilde{Q}+1}^Q \rho_q \sigma_{\alpha,q}^{2},  ~~~ & i = j;    \\
                          0, ~~~                                                    & i \neq j.
                      \end{cases}
        \end{align}
    \end{subequations}
    \textcolor{black}{The same finite-frame boundary convention as in Appendix B is adopted. The missing ISI contribution for the first OFDM symbol yields a diagonal boundary correction in the covariance matrix. Since this correction affects only one of the $M$ OFDM symbols, the correction is of order $\mathcal{O}(1/M)$ and is neglected to obtain the compact covariance form.}
    Similarly, the covariance of the ICI component $\hat{h}_{i}^{\mathrm{ICI}}$ is
    \begin{subequations}
        \begin{align}
              & \mathbb{E}\{ \hat{h}_{i}^{\mathrm{ICI}} (\hat{h}_{j}^{\mathrm{ICI}})^{\ast} \}                                                                                                               \\
            = & \! \sum_{q=\tilde{Q}+1}^{Q} \! \sigma_{\alpha,q}^{2}  \sum_{n_1, n_2}  \mathbb{E}\{ s_{n_1, m} s_{n,m}^{\ast} s_{n_2, m'}^{\ast} s_{n',m'} \} \phi_{n,n_1}^q (\phi_{n',n_2}^q)^{\ast} \notag \\
              & \times e^{-\jmath 2\pi (n_1 - n_2)\Delta_f \tau_q}    e^{\jmath 2 \pi (m-m') f_{\mathrm{d},q} T_{\mathrm{s}}}                                                                                \\
            = & \begin{cases}
                    \sum_{q=\tilde{Q}+1}^{Q}\sigma_{\alpha,q}^{2} \big( (\mu_4-1) \rho_q^2 + \rho_q \big), ~                & i=j;      \\
                    \sum_{q=\tilde{Q}+1}^{Q} \rho_q^2 \sigma_{\alpha,q}^{2} [\mathbf{a}_q]_{i} [\mathbf{a}_q^{\ast}]_{j}, ~ & i \neq j.
                \end{cases}
        \end{align}
    \end{subequations}
    The covariance of the noise component $\hat{h}_{i}^{\mathrm{awgn}}$ is given by
    \begin{subequations}
        \begin{align}
            \mathbb{E}\{ \hat{h}_{i}^{\mathrm{awgn}} (\hat{h}_{j}^{\mathrm{awgn}})^{\ast} \}
             & = \mathbb{E}\{z_{n,m} s_{n,m}^{\ast} z_{n',m'}^{\ast} s_{n',m'}\} \\
             & = \delta_{i,j} \sigma^2.
        \end{align}
    \end{subequations}

    Next, we derive the cross covariances between different components. The cross covariance between $\hat{h}_{i}^{\mathrm{free}}$ and $\hat{h}_{j}^{\mathrm{ISI}}$ is
    \begin{subequations}
        \begin{align}
            \mathbb{E}\{ \hat{h}_{i}^{\mathrm{free}} (\hat{h}_{j}^{\mathrm{ISI}})^{\ast} \}
             & = \!\! \sum_{q = \tilde{Q}+1}^Q \!\! \sigma_{\alpha,q}^{2} \!\sum_{n_1 = 0}^{N-1}\! \mathbb{E}\{ |s_{n,m}|^2 s_{n_1, m'-1}^{\ast} s_{n',m'} \}  \notag                                  \\
             & \hspace{-2.2 cm} \times (\phi_{n',n_1}^q)^{\ast} e^{\jmath 2 \pi \Delta_f ( n_1 (\tau_q - T_\mathrm{cp}) - n \tau_q )} e^{\jmath 2 \pi (m-m'+1) f_{\mathrm{d},q} T_{\mathrm{s}}} \!\!\! \\
             & = ~ 0.
        \end{align}
    \end{subequations}
    The cross covariance between $\hat{h}_{i}^{\mathrm{free}}$ and $\hat{h}_{j}^{\mathrm{ICI}}$ is given by
    \begin{subequations}
        \begin{align}
            \mathbb{E}\{ \hat{h}_{i}^{\mathrm{free}} (\hat{h}_{j}^{\mathrm{ICI}})^{\ast} \} =
              & \sum_{q = \tilde{Q}+1}^Q \sigma_{\alpha,q}^{2} \sum_{n_1 = 0}^{N-1} \mathbb{E}\{ |s_{n,m}|^2 s_{n_1, m'}^{\ast} s_{n',m'} \}  \notag               \\
              & \hspace{-1.5 cm} \times (\phi_{n',n_1}^q)^{\ast} e^{-\jmath 2 \pi (n-n_1) \Delta_f \tau_q} e^{\jmath 2 \pi (m-m') f_{\mathrm{d},q} T_{\mathrm{s}}} \\
            = & \begin{cases}
                    \sum_{q=\tilde{Q}+1}^{Q} \mu_4 \rho_q \sigma_{\alpha,q}^{2},                                        & i=j;      \\
                    \sum_{q=\tilde{Q}+1}^{Q} \rho_q \sigma_{\alpha,q}^{2} [\mathbf{a}_q]_{i} [\mathbf{a}_q^{\ast}]_{j}, & i \neq j.
                \end{cases}
        \end{align}
    \end{subequations}
    The cross covariance between $\hat{h}_{i}^{\mathrm{ISI}}$ and $\hat{h}_{j}^{\mathrm{ICI}}$ is
    \begin{subequations}
        \begin{align}
            \mathbb{E}\{ \hat{h}_{i}^{\mathrm{ISI}} (\hat{h}_{j}^{\mathrm{ICI}})^{\ast} \}
             & = \!\!\! \sum_{q = \tilde{Q}+1}^Q \!\!\! \sigma_{\alpha,q}^{2}  \!\!\sum_{n_1, n_2}\! \mathbb{E}\{\! s_{n_1, m-1} s_{n,m}^{\ast} s_{n_2, m'}^{\ast} s_{n',m'} \!\} \notag                           \\
             & \hspace{-2.2 cm} \times  \phi_{n,n_1}^q (\phi_{n',n_2}^q)^{\ast} e^{\jmath 2 \pi  \Delta_f (n_2 \tau_q - n_1 (\tau_q - T_{\mathrm{cp}}))} e^{\jmath 2 \pi (m-m'-1) f_{\mathrm{d},q} T_{\mathrm{s}}} \\
             & =  ~ 0.
        \end{align}
    \end{subequations}
    Finally, the cross covariance between the signal and noise components can be written as
    \begin{subequations}
        \begin{align}
              & \mathbb{E}\{ (\hat{h}_{i}^{\mathrm{free}} + \hat{h}_{i}^{\mathrm{ISI}} - \hat{h}_{i}^{\mathrm{ICI}}) (\hat{h}_{j}^{\mathrm{awgn}})^{\ast} \} \notag \\
            = & \mathbb{E}\{ (\hat{h}_{i}^{\mathrm{free}} + \hat{h}_{i}^{\mathrm{ISI}} - \hat{h}_{i}^{\mathrm{ICI}}) s_{n',m'} \} \mathbb{E}\{ z_{n',m'}^{\ast} \}  \\
            = & 0.
        \end{align}
    \end{subequations}

    Based on the above results, the $(i,j)$-th entry of the covariance matrix $\mathbf{R}$ can be written as
\begin{subequations}
    \begin{align}
        [\mathbf{R}]_{i,j} =
        \begin{cases}
            \sum_{q=1}^Q \tilde{\sigma}_{\alpha,q}^{2} + \sigma_{\mathrm{SL}}^2 ,                & i = j;    \\
            \sum_{q=1}^Q \tilde{\sigma}_{\alpha,q}^{2} [\mathbf{a}_q]_i [\mathbf{a}_q^{\ast}]_j, & i \neq j. \\
        \end{cases}
    \end{align}
\end{subequations}
Therefore, the covariance matrix can be expressed in compact form as
\begin{equation}
    \begin{aligned}
        \mathbf{R}
         & = \sum_{q=1}^Q \tilde{\sigma}_{\alpha,q}^{2} \mathbf{a}_q \mathbf{a}_q^H + \sigma_{\mathrm{SL}}^2 \mathbf{I}_{NM} \\
         & = \mathbf{A}_Q \boldsymbol{\Sigma}_{\alpha} \mathbf{A}_Q^H + \sigma_{\mathrm{SL}}^2 \mathbf{I}_{NM}.
    \end{aligned}
\end{equation}

\end{document}